\documentclass[journal]{IEEEtran}
\usepackage{cite}
\usepackage{amsmath}
\usepackage{amsfonts}
\usepackage{color}
\usepackage{amssymb}
\usepackage{array}
\usepackage{url}
\usepackage{graphicx}
\usepackage{subfigure}
\usepackage{longtable}
\usepackage{multirow}
\usepackage{marvosym}
\usepackage{textcomp}
\usepackage{threeparttable, tablefootnote}
\usepackage{enumerate}
\usepackage[font=small,labelsep=period]{caption}
\usepackage{dsfont}

\usepackage{xcolor}
\usepackage{soul}
\soulregister\cite7
\soulregister\ref7
\usepackage{algorithm,algpseudocode}
\makeatletter
\newcommand{\algmargin}{\the\ALG@thistlm}
\makeatother
\newlength{\whilewidth}
\algdef{SE}[parWHILE]{parWhile}{EndparWhile}[1]
  {\parbox[t]{\dimexpr\linewidth-\algmargin}{%
     \hangindent\whilewidth\strut\algorithmicwhile\ #1\ \algorithmicdo\strut}}{\algorithmicend\ \algorithmicwhile}%
\algnewcommand{\parState}[1]{\State%
  \parbox[t]{\dimexpr\linewidth-\algmargin}{\strut #1\strut}}

\algdef{SE}[parIF]{parIf}{EndparIf}[1]
  {\parbox[t]{\dimexpr\linewidth-\algmargin}{%
     \hangindent\ifwidth\strut\algorithmicif\ #1\ \algorithmicdo\strut}}{\algorithmicend\ \algorithmicif}%

\allowdisplaybreaks[4]
\newtheorem{theorem}{Theorem}

\ifCLASSINFOpdf
\else
\fi
\begin{document}
%
\title{Green Offloading in Fog-Assisted IoT Systems: An Online Perspective Integrating Learning and Control}
%
%
%

\author{Xin~Gao,~\IEEEmembership{Student Member,~IEEE,}
		Xi~Huang,~\IEEEmembership{Student Member,~IEEE,}
        Ziyu~Shao$^{*}$,~\IEEEmembership{Senior Member,~IEEE,}
        Yang~Yang,~\IEEEmembership{Fellow,~IEEE}                
\thanks{
X. Gao, X. Huang, Z. Shao and Y. Yang are with the School of Information Science and Technology, ShanghaiTech University, Shanghai 201210, China. (E-mail: \{gaoxin, huangxi, shaozy, yangyang\}@shanghaitech.edu.cn)
(*Corresponding author: Ziyu Shao)
}
}
\maketitle

\begin{abstract}
In fog-assisted IoT systems, it is a common practice to offload tasks from IoT devices to their nearby fog nodes to reduce task processing latencies and energy consumptions. However, the design of online energy-efficient scheme is still an open problem because of  various uncertainties in system dynamics such as processing capacities and transmission rates. Moreover, the decision-making process is constrained by resource limits on fog nodes and IoT devices, making the design even more complicated. In this paper, we formulate such a task offloading problem with unknown system dynamics as a combinatorial multi-armed bandit (CMAB) problem with long-term constraints on time-averaged energy consumptions. Through an effective integration of online learning and online control, we propose a \textit{Learning-Aided Green Offloading} (LAGO) scheme. In LAGO, we employ bandit learning methods to handle the exploitation-exploration tradeoff and utilize virtual queue techniques to deal with the long-term constraints. Our theoretical analysis shows that LAGO can reduce the average task latency with a tunable sublinear regret bound over a finite time horizon and satisfy the long-term time-averaged energy constraints. We conduct extensive simulations to verify such theoretical results.
\end{abstract}

\begin{IEEEkeywords}
Internet of Things, task offloading, energy consumption, fog computing, bandit learning, learning-aided control.
\end{IEEEkeywords}

\section{Introduction}

During recent years, the rapid development of Internet of Things (IoT) has spawned a wide range of applications\cite{lin2017survey}. 
In such applications, tasks are constantly generated on resource-constrained IoT devices\cite{chiang2016fog}. Such tasks need to be processed with high quality of service in terms of energy efficiency\cite{xiang2013energy}\cite{ge2015spatial} and task latency\cite{zhong2017heterogeneous}.
Therefore, it is a common practice to offload tasks from IoT devices to their nearby fog nodes with adequate processing capacities\cite{shi2016edge} via wireless connections.

By far, the design of online task offloading scheme is still an open problem.
There are three key challenges.
The \textit{first} challenge lies in how to reduce performance loss under various uncertainties in system dynamics. 
For example, available processing capacities and wireless channel states often vary among fog nodes and IoT devices, and these dynamics may change rapidly over time. 
In practice, these dynamics are hard to attain \textit{a priori} and their statistics can only be estimated from feedback information upon task completion.
Therefore, the decision making needs the aid of online learning to handle such uncertainties.
The \textit{second} challenge pertains to the \textit{exploration-exploitation tradeoff} in the course of online learning, which should be carefully addressed. This is because
1) over-exploitation, \textit{i.e.}, IoT devices stick to offloading tasks to particular fog nodes that have the empirically best performance,
may hinder the decision-making process from collecting more informative feedback from other potentially better fog nodes;
2) over-exploration, \textit{i.e.}, IoT devices blindly spread tasks among different fog nodes to gain new knowledge, may result in excessive latencies and energy consumptions.
The \textit{third} challenge is related to the constraints on energy consumptions due to resource limits on fog nodes and IoT devices.
Specifically, the advantage of offloading comes from the adequate processing capacities of fog nodes for task processing. 
Nonetheless, this benefit can be offset by considerable energy consumptions for wireless transmission. 
Such a non-trivial tradeoff requires a careful treatment to balance latency reduction and energy constraints.

\begin{table*}[!h]
\centering
\caption{Comparisons of related works}
\label{table: related works}
\begin{threeparttable}	
\begin{tabular}{|c|c|c|c|c|c|c|c|}
\hline
& MAB Model & Optimization Metrics & Constraints & Adopted Techniques & Online Learning & Online Control \\
\hline
\cite{sun2019adaptive} & Extended MAB & Latency & No & UCB-like technique & Yes & No \\
\hline
\cite{wang2019learning} & CMAB & Latency & No &  UCB-like technique & Yes & No \\
\hline
\cite{zhu2019blot} & Non-stationary MAB & Latency, energy & No & Discounted UCB & Yes & No \\
\hline
\cite{ghoorchian2019multi} & Budget-limited MAB & Latency, energy & Static & BPRPC-SWUCB & Yes & No \\
\hline
Our Work & Extended CMAB & Latency, energy & Time-averaged & UCB, virtual queue technique & Yes & Yes \\
\hline
\end{tabular}
\end{threeparttable}
\end{table*}

In this work, we address all the above challenges.
In particular, we focus on the task offloading between IoT devices and fog nodes in general fog-assisted IoT systems with unknown node processing capacities and link transmission rates.
The contributions and key results of this paper are summarized as follows.
\begin{itemize}
	\item[$\diamond$] \textbf{Problem Formulation:}
	We formulate the online task offloading problem to minimize the average task latency under long-term constraints on time-averaged energy consumptions.
	To deal with the uncertainties in system dynamics, we reformulate the problem as a constrained \textit{combinatorial multi-armed bandit} (CMAB) problem. Our model extends settings of the CSMAB-F model in \cite{li2019combinatorial} by allowing each arm to be played for multiple times during each time slot
	under the long-term time-averaged resource constraints.
		
	\item[$\diamond$] \textbf{Algorithm Design:} 
	To solve the problem, we propose an energy-efficient online task offloading scheme called \textit{LAGO (Learning-Aided Green Offloading)}.
	Through an integrated design, we adopt upper-confidence-bound1 (UCB1) method\cite{auer2002finite} to address the exploration-exploitation tradeoff in online learning, and employ Lyapunov optimization techniques\cite{neely2010stochastic} to make offloading decisions under long-term energy constraints.
	Besides, we also leverage other bandit learning methods to propose variants of LAGO.
	
	\item[$\diamond$] \textbf{Theoretical Analysis:} Our theoretical analysis shows that LAGO achieves a regret bound of $O(1/V+\sqrt{(\log{T})/T})$ over time horizon $T$ with tunable positive parameter $V$, subject to long-term constraints on energy consumptions. 
	The regret bound characterizes the reward loss incurred by both online control and online learning.
		
	\item[$\diamond$] \textbf{Numerical Evaluation:} We conduct extensive simulations to evaluate the performance of LAGO and its variants. Our simulation results show that our proposed schemes reduce task latency effectively under long-term time-averaged energy constraints.
\end{itemize}

The rest of this paper is organized as follows. We present our system model in Section \ref{sec: model} and problem formulation in Section \ref{sec: formulation}. 
Then we show our algorithm design in Section \ref{sec: algorithm} and performance analysis in Section \ref{sec: analysis}. 
Section \ref{sec: simulation} discusses our simulation results, while Section \ref{sec: conclusion} concludes this paper.

\section{Related Work}\label{sec: related work}

Regarding online offloading scheme design, existing works are generally carried out from two perspectives: online control and online learning.

\textbf{Online Control for Offloading:}
Works of this category formulated the online offloading problem as stochastic network optimization problems with different optimization metrics such as network throughput\cite{li2019dynamic}, total energy consumption\cite{pu2016d2d,gao2020pora,liu2019dynamic}, energy efficiency\cite{kim2018mobile}\cite{zhang2020dynamic}, and integrated metrics\cite{mao2016dynamic}\cite{cai2020jote}.
For example, 
Pu \textit{et al.}\cite{pu2016d2d} considered to minimize the total energy consumption while satisfying the energy budget constraint on each IoT device.
Kim \textit{et al.}\cite{kim2018mobile} aimed at maximizing the energy-utility efficiency of an IoT device under an energy consumption threshold and a minimum throughput guarantee. 
Mao \textit{et al.}\cite{mao2016dynamic} aimed to minimize the task latency and task dropping cost of an energy harvesting IoT device under deadline and battery discharging constraints.
Cai \textit{et al.}\cite{cai2020jote} jointly investigated the task and energy offloading in fog-enabled IoT systems with the goal of minimizing the weighted sum of task latency and energy consumption.
By adopting Lyapunov optimization techniques\cite{neely2010stochastic}, they transformed the problem into a series of subproblems and solved them on a per-time-slot basis.
However, such works implicitly assumed the instant system dynamics to be fully observable at the beginning of each time slot, which are hard to attain in practice. 

\textbf{Online Learning for Offloading:}
The other is online learning perspective. From such a perspective, existing works consider the online offloading problem in the cases with unknown system dynamics. To handle such uncertainties, they adopted various learning methods such as reinforcement learning\cite{chen2015energy}, neural networks\cite{chen2018optimized}\cite{min2019learning}, and bandit learning\cite{sun2019adaptive,wang2019learning,zhu2019blot,ghoorchian2019multi}. 
However, learning methods such as the Markov decision process (MDP) based reinforcement learning and neural network based deep learning often resort to heuristic designs and offline training processes with high computational complexities. 
In contrast, bandit learning provides a powerful framework for sequential decision-making problems under uncertainty with theoretical performance guarantee. It enjoys a wide adoption by a number of previous works. For example,
Sun \textit{et al.}\cite{sun2019adaptive} considered the task offloading among vehicles in vehicular edge computing systems with unknown offloading latencies. 
They proposed a distributed learning-based offloading scheme to minimize the average offloading latency in dynamic environments with time-varying offloading candidates. 
Besides, Wang \textit{et al.} \cite{wang2019learning} jointly considered task allocation and spectrum scheduling in the task offloading problem, with the aim to minimize task latencies under unknown computation resources of fog nodes. 
Zhu \textit{et al.}\cite{zhu2019blot} proposed an online task offloading scheme to minimize a combined cost of latency and energy consumption. 
Ghoorchian \textit{et al.}\cite{ghoorchian2019multi} aimed to minimize offloading latencies under static constraints on accumulated energy consumptions over a finite time horizon.

Different from previous works, we integrate online control and bandit learning 
to learn the unknown system dynamics effectively under long-term time-averaged energy consumption constraints. Such constraints ensure the energy consumption on each node below some given budget stochastically over a long period of time. 
A detailed comparison of our work with the bandit based works is shown in TABLE \ref{table: related works}. 

\textbf{Learning-aided Control under Bandit Settings:}
Bandit learning has also been applied in many other real world applications such as news article recommendation in personalized web services\cite{li2010contextual}, adaptive shortest-path routing in wireless networks\cite{liu2012adaptive}, sniffer channel assignment in cognitive radio networks\cite{xu2015sniffer}, \textit{etc}. The wide adoption of bandit learning in such real scenarios has motivated the development of diverse bandit learning frameworks, \textit{e.g.}, combinatorial bandits\cite{cesa2012combinatorial}, adversarial bandits\cite{auer1995gambling}, and contextual bandits\cite{tewari2017ads}. 
Among these frameworks, the most relevant to to our work is the combinatorial multi-armed bandit (CMAB). Recently, Li \textit{et al.}\cite{li2019combinatorial} proposed a novel CMAB model named CSMAB-F to ensure the fairness of arm selection besides reward maximization.
By extending the settings of CSMAB-F, we model our offloading problem into a constrained CMAB problem to deal with the system uncertainties under energy consumption constraints. The extensions lie in three aspects. First, we allow each arm to be selected repeatedly during the same time slot. Second, the reward of each arm is a function of the unknown system dynamics. Third, we consider the energy consumption constraints instead of fairness constraints.

\section{System Model}\label{sec: model}


\subsection{Basic Model}

We consider the interplay between one IoT device and $N$ fog nodes in a time-slotted fog-assisted IoT system.\footnote{Note that we do not consider multi-hop offloading, \textit{i.e.}, the offloading with tasks transferred among fog nodes.}
We assume that the IoT device has potential connections to $N$ fog nodes. We use $\mathcal{N}=\{0,1,\dots,N\}$ to denote the set of all nodes, in which the IoT device is indexed by $0$ and the indices of fog nodes vary from $1$ to $N$.
During each time slot $t$, due to wireless channel state variations, 
the IoT device may have access to only a subset of the fog nodes.
Accordingly, we denote the set of nodes accessible to the IoT device (including itself) by $\mathcal{N}(t) \subseteq \mathcal{N}$.
The system operates through the following three stages during each time slot $t$.

\textbf{Stage 1:} \textit{Task Generation.}
At the beginning of time slot $t$, the IoT device generates a number of new tasks, denoted by set $\mathcal{A}(t)$. 
The task generation process is assumed \textit{i.i.d.} across time slots such that $|\mathcal{A}(t)| \leq a_{\text{max}}$ for some constant $a_{\text{max}}$. 
Each task $i\in\mathcal{A}(t)$ has a size of $L_{i}(t)$ bits and a computation demand of $W_{i}(t)$ CPU cycles.
The values of $L_{i}(t)$ and $W_{i}(t)$ are known to the IoT device upon task arrival and upper bounded by constants $l_{\text{max}}$ and $w_{\text{max}}$, respectively.

\textbf{Stage 2:} \textit{Offloading Decision Making.}
For each new task, the IoT device should decide whether to offload it and which node it is offloaded to. 
We denote the \textit{offloading decision} for each task $i\in\mathcal{A}(t)$ by $I_{i}(t) \in \mathcal{N}(t)$. 
Particularly, $I_{i}(t)=0$ indicates that task $i$ will be processed locally on the IoT device; otherwise, task $i$ will be offloaded to fog node $I_{i}(t)$.
For ease of analysis, we assume that each new task can be either processed locally or uploaded to one of the fog nodes. 
Nonetheless, our model can be directly extended to the scenarios with splittable tasks. 

\textbf{Stage 3:} \textit{Task Scheduling.}
Given offloading decisions, tasks are scheduled and processed accordingly. 
At the end of time slot $t$, the processing results together with the metrics of task latencies are sent back to the IoT device.


\subsection{Optimization Objectives}


\textbf{Task Latency:}
Task latency is one of the key QoS measurements in many IoT applications\cite{shi2016edge}.
For each task, its latency mainly consists of the \textit{transmission latency} (if it is offloaded to one of the fog nodes) and the \textit{processing latency}.
Below we define related notations of such metrics in detail.

\textit{Transmission Latency:}
We denote the transmission latency of each task $i\in\mathcal{A}(t)$ by $D_{i,I_{i}(t)}^{(tr)}(t)$.
If $I_{i}(t) = 0$, the task will be processed locally
and no transmission latency will be incurred ($D_{i,I_{i}(t)}^{(tr)}(t) = 0$).
Otherwise, task $i$ will be offloaded to fog node $I_{i}(t)$. 
By defining $R_{i,n}(t)$ as the transmission rate allocated to offload task $i$ to fog node $n$, we have
\begin{equation}\label{definition: transmission latency}
	D_{i,I_{i}(t)}^{(tr)}(t)=L_{i}(t)\mathds{1}\{ I_{i}(t)>0\} / 
	R_{i,I_{i}(t)}(t),
\end{equation}
in which we note that $L_{i}(t)$ denotes the size of task $i$ and $\mathds{1}\{\cdot\}$ is the indicator function.

\textit{Processing Latency:}
We denote the processing latency of task $i\in\mathcal{A}(t)$ by $D_{i,I_{i}(t)}^{(pr)}(t)$. 
Recall that the computation demand $W_{i}(t)$ of task $i$ is in the units of CPU cycles. 
Let $F_{i,n}(t)$ be the CPU cycle frequency allocated to task $i$ by node $n$, the processing latency of task $i$ is
\begin{equation}\label{definition: processing latency}
	D_{i,I_{i}(t)}^{\left(pr\right)}\left(t\right)
	=
	W_{i}\left(t\right)/F_{i,I_{i}\left(t\right)}\left(t\right).
\end{equation}
By (\ref{definition: transmission latency}) and (\ref{definition: processing latency}), the total latency of task $i$ is given by
\begin{equation}	\label{eq: task latency}
\begin{split}
	&D_{i,I_{i}(t)}(t)=
	D_{i,I_{i}(t)}^{(tr)}(t)+D_{i,I_{i}(t)}^{(pr)}(t)\\ 
	&=L_{i}(t)\mathds{1} \{ I_{i}(t)>0\} /R_{i,I_{i}(t)}(t)+W_{i}(t)/F_{i,I_{i}(t)}(t).	
\end{split}
\end{equation}

In practice, due to wireless channel variations, the transmission rate $R_{i, n}\left(t\right)$ is hard to obtain before task $i$ is assigned,
and the allocated CPU cycle frequency $F_{i, n}(t)$ is revealed only after task $i$'s completion. 
To handle such uncertainties, we treat both $R_{i,n}\left(t\right)$ and $F_{i,n}(t)$ as random variables with unknown distributions and means.
We assume that both $R_{i,n}\left(t\right)$ and $F_{i,n}(t)$ are \textit{i.i.d.} across different tasks and lower bounded by constants $r_{\text{min}}$ and $f_{\text{min}}$, respectively.
In addition, we assume the existence of the mean of their reciprocals, denoted by 
$\rho_{n}\triangleq\mathbb{E}\left[1/R_{i,n}\left(t\right)\right]$
and
$\phi_{n}\triangleq \mathbb{E}\left[1/F_{i,n}\left(t\right)\right]$, respectively.
Note that $\rho_{n}$ can be viewed as the average transmission latency per bit to node $n$, and $\phi_{n}$ can be viewed as the average task processing latency per CPU cycle on node $n$.

\textbf{Energy Consumption:} 
Due to resource limits on the IoT device and fog nodes, energy efficiency is also considered in our work.
For the IoT device, its energy consumption is mainly incurred by wireless transmission and local task processing. 
For each fog node, its energy consumption is mainly caused by CPU processing.
For each time slot $t$, we use $\eta_{n}(t)$ to denote the energy consumption of sending one bit to fog node $n\in\mathcal{N}(t)\!\setminus\!\{0\}$, 
and $\kappa_{n}(t)$ to denote the energy consumption for task processing per CPU cycle on node $n\in\mathcal{N}(t)$.
We assume that $\eta_{n}(t)$ and $\kappa_{n}(t)$ are upper bounded by some positive constants $\eta_{\text{max}}$ and $\kappa_{\text{max}}$, respectively. Moreover, we assume the availability of $\eta_{n}(t)$ and $\kappa_{n}(t)$ at the beginning of time slot $t$.
Accordingly, the total energy consumptions on the IoT device and each fog node $n$ are respectively given as follows:
\begin{align}\label{eq: user energy}
	E_{0}(t)=&\sum_{i\in\mathcal{A}(t)}\kappa_{0}(t)W_{i}(t)\mathds{1}\{ I_{i}(t)=0\} \nonumber \\
	&+\sum_{i\in\mathcal{A}(t)}\sum_{n=1}^{N}\eta_{n}(t)L_{i}(t)\mathds{1} \{ I_{i}(t)=n\},\\
	E_{n}(t)=&\sum_{i\in\mathcal{A}(t)}\kappa_{n}(t)W_{i}(t)\mathds{1}\{ I_{i}(t)=n\}.\label{eq: fog energy}
\end{align}

Considering that each node $n \in \mathcal{N}$ often has a limited energy capacity, its long-term time-averaged energy consumption needs to be ensured within a given budget $b_{n}$, \textit{i.e.},
\begin{equation}\label{ineq: energy constraint}
	\limsup_{t\rightarrow\infty}\frac{1}{t}\sum_{\tau=0}^{t-1}\mathbb{E}\left[E_{n}\left(\tau\right)\right]\leq b_{n},\forall n\in\mathcal{N}.
\end{equation} 


\section{Problem Formulation}\label{sec: formulation}

Based on the system model presented in Section \ref{sec: model}, our problem formulation is given as follows:
\begin{equation}\label{p: initial}
\begin{array}{cl}
	\underset{\{I_{i}(t)\}_{i,t}}{\text{minimize}}
	&\displaystyle \sum_{t=0}^{T-1}\sum_{i\in\mathcal{A}\left(t\right)}\mathbb{E}\left[D_{i,I_{i}\left(t\right)}\left(t\right)\right]\\
	\text{subject to}
	&\displaystyle (\ref{ineq: energy constraint}), I_{i}\left(t\right)\in \mathcal{N}(t),\forall i\in\mathcal{A}(t),\forall t.
\end{array}
\end{equation}
Problem (\ref{p: initial}) aims to minimize the expected total task latency over finite time horizon $T$ under long-term time-averaged energy consumption constraints.
Note that when the allocated CPU cycle frequencies $\{F_{i,n}(t)\}_{i,n}$ and transmission rates $\{R_{i,n}(t)\}_{i,n}$ are given at the beginning of each time slot $t$, problem (\ref{p: initial}) can be solved asymptotically optimally by Lyapunov optimization techniques\cite{neely2010stochastic}. 
In our work, we consider the more general scenario when such information is unknown \textit{a priori}.
Therefore, it is necessary to estimate these uncertainties to facilitate the decision-making process. Particularly, in each time slot, the IoT device should update its estimates about the means of $\{F_{i,n}(t)\}_{i,n}$ and $\{R_{i,n}(t)\}_{i,n}$ based on the latest observation of task latencies, and then make effective offloading decisions based on the estimated values.

We find that our problem can be modeled under the multi-armed bandit (MAB)\cite{bubeck2012regret} framework, since it is a sequential decision problem under uncertainty.
There is a wide range of different MAB models\cite{lattimore2019bandit}, and the one that most fits our problem is the combinatorial multi-armed bandit (CMAB) model.
In the basic CMAB model, a player interacts with an environment over a finite number of rounds. In each round, the player chooses multiple arms (actions) among a set of candidate arms to play. Then the environment reveals a reward to the player for each played arm. The reward of each arm is a random variable which follows some unknown distribution and is \textit{i.i.d.} across different rounds. The goal of the player is to maximize the total received reward.
The CSMAB-F model proposed in \cite{li2019combinatorial} extends the settings of the canonical CMAB model by assuming that each arm could sometimes be unavailable and considering the fairness of arm selection.
To adapt such a model to our problem, we further extend its settings and reformulate problem (\ref{p: initial}) as a constrained CMAB problem. Now we discuss details of the reformulation in the following part of the section.

\textbf{Problem Reformulation:}
By regarding the IoT device as the player and each offloading choice as an arm, we reformulate problem (\ref{p: initial}) as a CMAB problem with long-term time-averaged energy constraints. Our extensions to the settings of the CSMAB-F model are as follows.
First, we assume that each arm can be played for multiple times during each time slot.
Second, the reward of each arm is a linear function of the uncertainties to be estimated, rather than the uncertainties themself.
Third, we consider the constraints on the long-term time-averaged energy consumptions, which are more complex than the fairness constraints.
Under such settings, there are $(N+1)$ arms in total since tasks can also be processed locally by the IoT device.
In each time slot $t$, the player chooses an arm $I_{i}(t)$ from subset $\mathcal{N}(t)$ for each new task $i\in\mathcal{A}(t)$, thus the super arm being chosen during time slot $t$ is $\{I_{i}(t)\}_{i\in\mathcal{A}(t)}$.
If choosing arm $n$ for task $i$, the player will receive a random reward of
\begin{equation}\label{def: task reward}
X_{i,I_{i}(t)}(t)\triangleq -D_{i,I_{i}(t)}(t).
\end{equation}
Then the player's goal is to maximize the expected time-averaged reward of $T$ time slots under constraints (\ref{ineq: energy constraint}).

We use $X^{*}$ to denote the maximal expected time-averaged reward achieved by the optimal policy. 
Then given offloading decision $\{I_{i}(t)\}_{i,t}$, the regret over $T$ time slots is given by
\begin{equation}\label{def: regret}
	R\left(T\right)\triangleq X^{*}-\frac{1}{T}\sum_{t=0}^{T-1}\sum_{i\in\mathcal{A}\left(t\right)}\mathbb{E}\left[X_{i,I_{i}\left(t\right)}\left(t\right)\right].
\end{equation}
Note that maximizing the reward is equivalent to minimizing the regret since $X^{*}$ is a constant. Therefore, solving problem (\ref{p: initial}) is equivalent to finding an optimal solution for the following problem:
\begin{equation}\label{p: original}
\begin{array}{cl}
	\underset{\{I_{i}(t)\}_{i,t}}{\text{minimize}}
	&\displaystyle R(T)\\
	\text{subject to}
	&\displaystyle (\ref{ineq: energy constraint}), I_{i}\left(t\right)\in \mathcal{N}(t),\forall i\in\mathcal{A}(t),\forall t.
\end{array}
\end{equation}
By definitions (\ref{def: task reward}) and (\ref{def: regret}), we have
\begin{equation}\label{eq: regret with task latency}
	R\left(T\right)=X^{*}+\frac{1}{T}\sum_{t=0}^{T-1}\sum_{i\in\mathcal{A}\left(t\right)}\mathbb{E}\left[D_{i,I_{i}\left(t\right)}\left(t\right)\right].
\end{equation}
Substituting the task latency $D_{i,I_{i}(t)}(t)$ in (\ref{eq: regret with task latency}) with (\ref{eq: task latency}) gives
\begin{equation}\label{eq: regret with mean}
\begin{split}
	R(T)=X^{*}&\!+\!\frac{1}{T}\sum_{t=0}^{T-1}\sum_{i\in\mathcal{A}(t)}\mathbb{E}\!\left[\rho_{I_{i}(t)}L_{i}(t)\mathds{1}\{ I_{i}(t)\!>\!0\}\right] \\
	&+\frac{1}{T}\sum_{t=0}^{T-1}\sum_{i\in\mathcal{A}\left(t\right)}\mathbb{E}\left[\phi_{I_{i}(t)}W_{i}(t)\right],
\end{split}
\end{equation}
where $\rho_{n}\triangleq \mathbb{E}\left[1/R_{i,n}\left(t\right)\right]$ and $\phi_{n}\triangleq \mathbb{E}\left[1/F_{i,n}\left(t\right)\right]$.

To solve problem (\ref{p: original}), there are two challenges. 
One is to estimate the unknown parameters $\{\rho_{n}\}_{n\in\mathcal{N}\setminus\{0\}}$ and $\{\phi_{n}\}_{n\in\mathcal{N}}$ in the objective function (\ref{eq: regret with mean}). 
The other is to satisfy the long-term time-averaged energy consumption constraints.


\section{Algorithm Design}\label{sec: algorithm}


Motivated by the idea of integrating bandit learning and virtual queue techniques in the recent work\cite{li2019combinatorial},
we propose a Learning-Aided Green Offloading (LAGO) scheme to solve problem (\ref{p: original}).
The pseudocode of LAGO is shown in Algorithm \ref{alg: policy}.
In the following, we demonstrate the design of LAGO in detail.

\subsection{Online Learning with UCB1 Method}

To achieve efficient online learning, we need to address the tradeoff between leveraging current knowledge (exploitation) and acquiring new knowledge (exploration), \textit{i.e.}, exploitation-exploration dilemma.
In our problem formulation, the unit transmission latency $\rho_{n}$ ($n>0$) and unit processing latency $\phi_{n}$ can be learned based on the reward received after arm $n$ has been pulled. 
The more times arm $n$ has been played for, the more reliable the estimates of $\rho_{n}$ ($n>0$) and $\phi_{n}$ are. 
However, if improperly conducted, such online learning may give rise to a higher regret when arm $n$ is suboptimal. 

To address such a tradeoff, we adopt UCB1\cite{auer2002finite} in the design of LAGO.
We denote $h_{n}(t)$ as the number of times that arm $n$ is selected during the first $t$ time slots. Moreover, we denote $\bar{\rho}_{n}(t)$ and $\bar{\phi}_{n}(t)$ as the empirical means of $1/R_{i,n}(t)$ and $1/F_{i,n}(t)$, respectively. These definitions are shown as follows:
\begin{align}
	&h_{n}\left(t\right)\triangleq \sum_{\tau=0}^{t-1}\sum_{i\in\mathcal{A}\left(\tau\right)}\mathds{1}\left\{ I_{i}\left(\tau\right)=n\right\},\forall n\in\mathcal{N},\label{eq: play number} \\
	&\bar{\rho}_{n}(t)\!\triangleq\! \frac{1}{h_{n}(t)}\!\sum_{\tau=0}^{t-1}\!\sum_{i\in\mathcal{A}(\tau)}\!\frac{\mathds{1}\{ I_{i}(\tau)=n\}}{R_{i,n}(\tau)},\forall n\in\mathcal{N}\setminus\{0\},\label{eq: empirical mean about transmit} \\
	&\bar{\phi}_{n}\left(t\right)\triangleq \frac{1}{h_{n}\left(t\right)}\sum_{\tau=0}^{t-1}\sum_{i\in\mathcal{A}\left(\tau\right)}\frac{\mathds{1}\left\{ I_{i}\left(\tau\right)=n\right\}}{F_{i,n}\left(\tau\right)},\forall n\in\mathcal{N}. \label{eq: empirical mean about offload}
\end{align}
Then the UCB1 estimates of $\rho_{n}$ and $\phi_{n}$ are updated during each time slot $t$ as follows:
\begin{align}
	&\hat{\rho}_{n}(t)\!=\![\bar{\rho}_{n}(t)\!-\!\rho_{\text{max}}\!\sqrt{3\log t / (2h_{n}(t))}]^{+},\!\forall n\!\in\mathcal{N}\!\setminus\!\{\! 0\!\}, \label{eq: UCB estimate about transmit} \\
	&\hat{\phi}_{n}(t)\!=\![\bar{\phi}_{n}(t)\!-\!\phi_{\text{max}}\!\sqrt{3\log t/(2h_{n}(t))}]^{+},\forall n\in\mathcal{N}, \label{eq: UCB estimate about process}
\end{align}
in which $[\cdot]^{+}\triangleq \max\{\cdot, 0\}$, $\rho_{\text{max}}\triangleq 1/r_{\text{min}}$, $\phi_{\text{max}}\triangleq 1/f_{\text{min}}$.

In (\ref{eq: UCB estimate about transmit}) and (\ref{eq: UCB estimate about process}), the terms $\rho_{\text{max}}\sqrt{3\log t / (2h_{n}(t))}$ and $\phi_{\text{max}}\sqrt{3\log t / (2h_{n}(t))}$ are called the \textit{confidence radius}\cite{slivkins2019introduction}. The confidence radius represents the ``optimism'' or the degree of uncertainty for selecting the arm.
If an arm has been played for a sufficient number of times, then the uncertainty in its estimates will be small, resulting in a small confidence radius.
In contrast, if it has rarely been explored, then the UCB1 method tends to optimistically over-estimate the values of $\rho_n$ and $\phi_n$ by a large confidence radius.
In particular, for each arm $n$, its update equations with respect to $h_{n}(t)$, $\bar{\rho}_{n}(t)$ and $\bar{\phi}_{n}(t)$ are given as follows:
\begin{align}
	& h_{n}\left(t\right)=h_{n}\left(t-1\right)+\sum_{i\in\mathcal{A}\left(t\right)}\mathds{1}\left\{ I_{i}\left(t\right)=n\right\},\forall n\in\mathcal{N}, \label{eq: counter update} \\
	& \bar{\rho}_{n}\left(t\right)=\bar{\rho}_{n}(t-1)h_{n}(t-1)/h_{n}(t) \nonumber \\
	&+\!\sum_{i\in\mathcal{A}(t)}\mathds{1}\{ I_{i}(t)=n\}/(R_{i,n}(t)h_{n}(t)),\forall n\in\mathcal{N}\setminus\{0\}, \label{eq: transmit update} \\
	& \bar{\phi}_{n}(t)=\bar{\phi}_{n}(t-1)h_{n}(t-1)/h_{n}(t) \nonumber \\
	&~~~~~~~+\sum_{i\in\mathcal{A}(t)}\mathds{1}\{ I_{i}(t)=n\}/(F_{i,n}(t)h_{n}(t)),\forall n\in\mathcal{N}. \label{eq: process update}
\end{align}
Note that updating $\bar{\rho}_{n}(t)$ and $\bar{\phi}_{n}(t)$ requires the exact values of $\{R_{i,I_{i}(t)}(t)\}_{i\in\mathcal{A}(t),I_{i}(t)>0}$ and $\{F_{i,I_{i}(t)}(t)\}_{i\in\mathcal{A}(t)}$. These values can be obtained from the feedback received by the IoT device at the end of each time slot $t$. Specifically, they are given by
\begin{align}
	&R_{i,I_{i}(t)}(t)=L_{i}(t)/d_{i}^{(tr)}(t),\forall i\in\mathcal{A}(t), I_{i}(t)>0, \\
	&F_{i,I_{i}(t)}(t)=W_{i}(t)/d_{i}^{(pr)}(t),\forall i\in\mathcal{A}(t),
\end{align}
where $d_{i}^{(tr)}(t)$ and $d_{i}^{(pr)}(t)$ are transmission latency and processing latency of task $i\in\mathcal{A}(t)$, respectively.
Other bandit learning methods such as UCB variants\cite{auer2002finite} and $\epsilon$-greedy\cite{vermorel2005multi} can also be used in our framework. 
We consider such variants in our simulations, which will be specified in Section \ref{sec: simulation}.

\subsection{Energy Budget Guarantee with Virtual Queue Technique}

\begin{figure}[!t]
	\centering
	\includegraphics[scale=0.4]{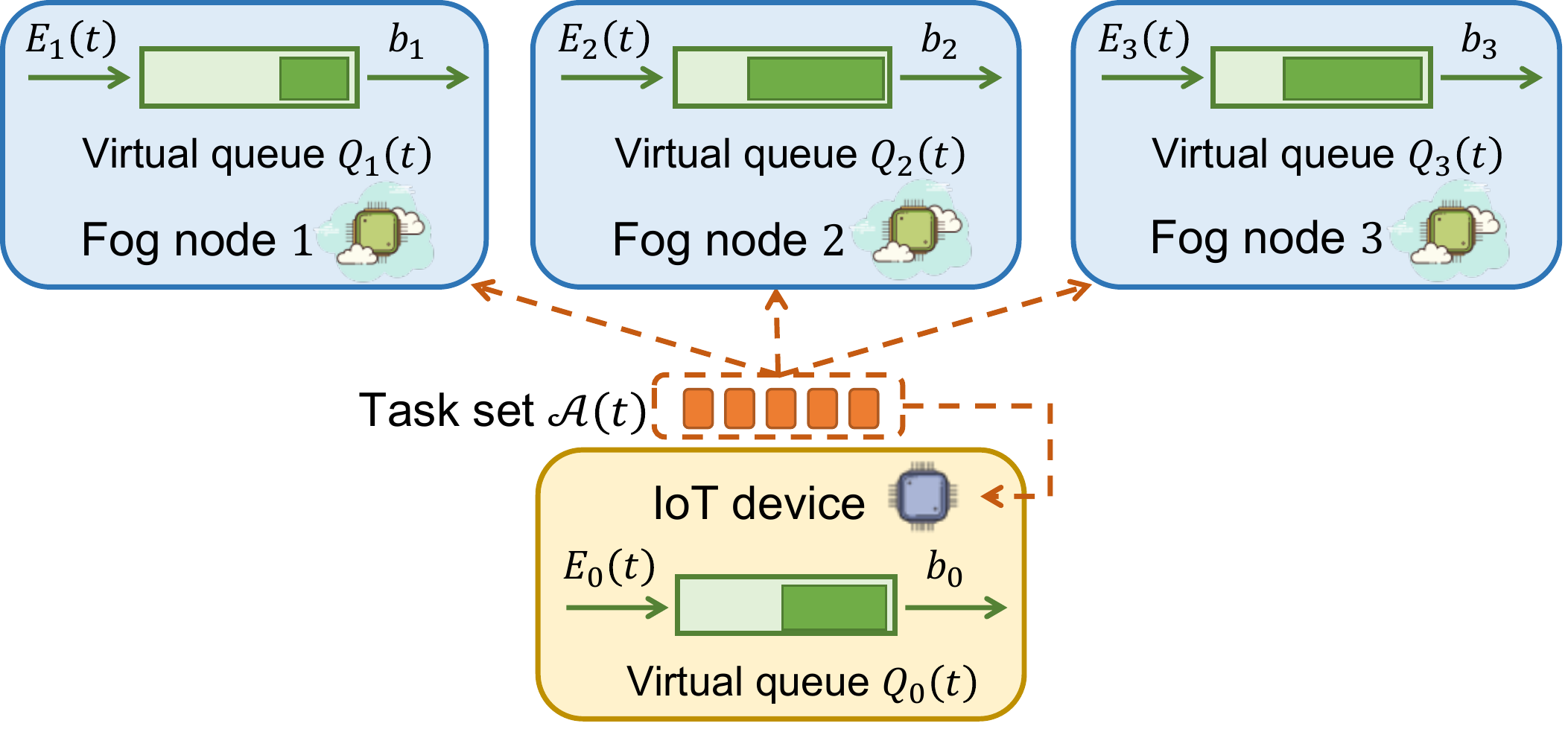}
	\caption{An illustration of virtual queues in a fog-assisted IoT system with one IoT device and three fog nodes. Each node $n\in\{0,1,2,3\}$ maintains a virtual queue $Q_{n}(t)$ with an input of $E_{n}(t)$ and an output of $b_{n}$ during each time slot $t$. If the queueing process $\{Q_{n}(t)\}_{t}$ is strongly stable, then the long-term time-averaged energy consumption on node $n$ can be guaranteed under the budget $b_{n}$.}
	\label{fig: virtual queue}
\end{figure} 

To handle the long-term constraints (\ref{ineq: energy constraint}) on time-averaged energy consumptions, we apply the idea of transforming time-averaged constraints into queue stability problems\cite{neely2010stochastic}. Specifically, we first introduce a virtual queue $Q_{n}(t)$ for each node $n\in\mathcal{N}$ with $Q_{n}(0)=0$.
Figure \ref{fig: virtual queue} demonstrates such virtual queues by an example.
During each time slot $t$, the output of queue $Q_{n}(t)$ is $b_{n}$, \textit{i.e.}, the energy budget of node $n$.
The input of queue $Q_{n}(t)$ is $E_{n}(t)$, \textit{i.e.}, the energy consumption on node $n$ during time slot $t$. 
The backlog size of queue $Q_{n}(t)$ is updated as follows:
\begin{equation}\label{eq: queue update}
	Q_{n}\left(t+1\right)=\left[ Q_{n}\left(t\right)-b_{n}\right]^{+} +E_{n}\left(t\right).
\end{equation}
According to \cite{neely2010stochastic}, if LAGO can guarantee the stability of the queueing process $\{Q_{n}(t)\}_t$, then the energy consumption constraint on node $n$ can also be ensured.

In every time slot $t$, under LAGO, the IoT device makes offloading decisions based on virtual queue backlog sizes, task properties, and UCB1 estimates introduced in (\ref{eq: UCB estimate about transmit}) and (\ref{eq: UCB estimate about process}).
We define price functions $\{v_{i,n}(t, \rho, \phi)\}_{n\in\mathcal{N}}$ for each task $i\in\mathcal{A}(t)$ to represent the price of allocating task $i$ to node $n$:
\begin{equation}\label{def: auxiliary func}
\begin{split}
	&v_{i,n}\left(t,\rho,\phi\right)\triangleq \big(Q_{n}\left(t\right)\kappa_{n}\left(t\right)W_{i}\left(t\right)\\
	&~~~~~~~+Q_{0}(t)\eta_{n}(t)L_{i}\left(t\right)\mathds{1}\left\{ n>0\right\}\big) \\
	&~~~~~~~+V\left(\phi W_{i}\left(t\right)+\rho L_{i}\left(t\right)\mathds{1}\left\{ n>0\right\} \right)
	,\forall n\in\mathcal{N},
\end{split}
\end{equation}
where $V$ is a tunable positive parameter. 
For each task $i\in\mathcal{A}(t)$, the IoT device chooses the node with the minimum price under estimates $\{\hat{\rho}_{n}(t)\}_{n\in\mathcal{N}(t)\setminus\{0\}}$ and $\{\hat{\phi}_{n}(t)\}_{n\in\mathcal{N}(t)}$, \textit{i.e.},
\begin{equation}\label{eq: our policy}
	I_{i}(t)=\arg\min_{n\in\mathcal{N}(t)}v_{i,n}(t,\hat{\rho}_{n}(t),\hat{\phi}_{n}(t)).
\end{equation}
Regarding such a node selection policy, we have the following remarks.

\textit{Remark 1:}
The tunable parameter $V$ in the price (\ref{def: auxiliary func}) measures the relative importance of minimizing task latency and satisfying energy constraints. Specifically, the price $v_{i,n}(t,\hat{\rho}_{n}(t),\hat{\phi}_{n}(t))$ is the weighted sum of two major terms. The first term $Q_{n}(t)\kappa_{n}(t)W_{i}(t)+Q_{0}(t)\eta_{n}(t)L_{i}(t)\mathds{1}\{ n>0\}$ is the weighted sum of the processing and wireless transmission energy consumption when allocating task $i$ to node $n$. The second term $\hat{\phi}_{n}(t) W_{i}(t)+\hat{\rho}_{n}(t)\mathds{1}\{ n>0\}$ is the estimated latency of task $i$ when it is allocated to node $n$. 
When the value of $V$ is large, the second term becomes dominant in the price $v_{i,n}(t,\hat{\rho}_{n}(t),\hat{\phi}_{n}(t))$. As a result, under LAGO, the IoT device tends to choose the node with the minimal estimated latency. 
To the contrary, when the value of $V$ is small, the first term becomes dominant and LAGO tends to select the node that incurs low energy consumptions of processing and wireless transmission.

\textit{Remark 2:}
To ensure the energy constraints in (\ref{ineq: energy constraint}), LAGO maintains a virtual queue for each node to track the ``debt'' to selecting the node.
If the time-averaged energy consumption on node $n$ exceeds the budget $b_{n}$, its associated ``debt'' (\textit{i.e.}, virtual queue backlog size $Q_{n}(t)$) will become large. As a result, the price $v_{i,n}(t,\hat{\rho}_{n}(t),\hat{\phi}_{n}(t))$ of selecting node $n$ for each task $i$ becomes high and node $n$ is unlikely to be selected under LAGO.
Moreover, if $Q_{0}(t)$ is small when compared with other virtual queue backlog sizes $\{Q_{n}(t)\}_{n\in\mathcal{N}(t)\setminus\{0\}}$, the IoT device is more willing to process tasks locally to help reduce the energy consumptions on fog nodes.

\begin{algorithm}[!t]
\caption{Learning-Aided Green Offloading (LAGO)}
\label{alg: policy}
\begin{algorithmic}[1]
  \State Initialize $\bar{\rho}_{n}(0)=\hat{\rho}_{n}(0)=0$ for each fog node $n\in\mathcal{N}\setminus\{0\}$, and $\bar{\phi}_{n}(0)=\hat{\phi}_{n}(0)=h_{n}(0)=0$ for each node $n\in\mathcal{N}$. Initialize $V=1$.
  In each time slot $t\in\{0,1,\dots\}$:
    \Statex \begin{center}
    	\textbf{\textit{Online Learning}}
    \end{center}
    \For {each node $n\in\mathcal{N}(t)$}
      \If {$h_{n}(t)>0$}
        \State Update $\hat{\phi}_{n}(t)$ according to (\ref{eq: UCB estimate about process}).\label{algline: estimate 1}
        \State Update $\hat{\rho}_{n}(t)$ according to (\ref{eq: UCB estimate about transmit}) if $n$ is a fog node.\label{algline: estimate 2}
      \EndIf
    \EndFor
    \Statex \begin{center}
    	\textbf{\textit{Task Offloading}}
    \end{center}
    \For {each task $i\in\mathcal{A}(t)$} \label{algline: task offloading begin}
      \parState {Set $I_{i}\left(t\right)\leftarrow \arg\min_{n\in\mathcal{N}\left(t\right)}v_{i,n}(t, \hat{\rho}_{n}(t),\hat{\phi}_{n}(t))$ and assign task $i$ to node $I_{i}(t)$.} \label{algline: node selection}
    \EndFor \label{algline: task offloading end}
    \parState {Update virtual queues $\{Q_{n}(t)\}_{n\in\mathcal{N}(t)}$ according to (\ref{eq: queue update}).}
    \Statex \begin{center}
    	\textbf{\textit{Update of Selection Counts and Empirical Means}}
    \end{center}
    \For {each node $n\in\mathcal{N}(t)$}
      \State Update $h_{n}(t)$ and $\bar{\phi}_{n}(t)$ according to (\ref{eq: counter update}) and (\ref{eq: process update}).
      \State Update $\bar{\rho}_{n}(t)$ according to (\ref{eq: transmit update}) if $n$ is a fog node.
    \EndFor
\end{algorithmic}
\end{algorithm} 

As shown in Algorithm \ref{alg: policy}, LAGO contains three stages in every time slot. First, it learns the UCB1 estimates of unknown $\phi_{n}$ and $\rho_{n}$. 
Then it makes offloading decisions for every new task based on UCB1 estimates. 
Finally, LAGO updates the counts of arm selections and empirical means based on task feedbacks during current time slot.
Note that the computational complexity of LAGO is $O(Na_{\text{max}})$, which is mainly from the task offloading process (lines
\ref{algline: task offloading begin}-\ref{algline: task offloading end}
in Algorithm \ref{alg: policy}).

\section{Theoretical Analysis}\label{sec: analysis}

In this section, we conduct theoretical analysis to investigate the performance of LAGO in terms of the upper bounds for virtual queue backlog sizes and its regret.

\subsection{Energy Consumption Bound}

For any energy budget vector $\boldsymbol{b}=(b_{0},\dots,b_{N})$, it is said to be \textit{feasible} if there exists an offloading policy such that all the long-term time-averaged energy constraints (\ref{ineq: energy constraint}) are satisfied. 
By defining the set of all feasible energy budget vectors as the \textit{maximal feasibility region} of the system, we have the following theorem.

\begin{theorem}\label{theorem: feasibility}
	Suppose that the energy budget vector $\boldsymbol{b}$ lies in the interior of the maximal feasibility region of the system. Then the long-term time-averaged energy constraints (\ref{ineq: energy constraint}) are satisfied under LAGO. 
	Moreover, the queueing processes (\ref{eq: queue update}) are strongly stable and there exists some positive constant $\epsilon$ such that
	\begin{equation}\label{ineq: queue upper bound}
		\limsup_{t\rightarrow\infty}\frac{1}{t}\sum_{\tau=0}^{t-1}\sum_{n\in\mathcal{N}}\mathbb{E}\left[Q_{n}\left(\tau\right)\right]\leq\frac{B+V\left(\theta_{1}+\theta_{2}\right)}{\epsilon},
	\end{equation}
	where $B\!\triangleq\!\sum_{n\in\mathcal{N}}b_{n}^{2}/2+a_{\text{max}}^{2}(\max\{ \kappa_{\text{max}}^{2}w_{\text{max}}^{2},\eta_{\text{max}}^{2}l_{\text{max}}^{2}\}/2 +N^{2}a_{\text{max}}^{2}\kappa_{\text{max}}^{2}w_{\text{max}}^{2}/2$, $\theta_{1}\!\triangleq\! 2w_{\text{max}}\phi_{\text{max}}a_{\text{max}}$, $\theta_{2}\!\triangleq \!2l_{\text{max}}\rho_{\text{max}}a_{\text{max}}$.
\end{theorem}

We adopt Lyapunov drift analysis\cite{neely2010stochastic} to prove Theorem \ref{theorem: feasibility}.
First, we introduce the notion of Lyapunov drift to characterize the change of the total virtual queue backlog size between successive time slots. 
Next, we focus on the drift-plus-regret, which is the weighted sum of Lyapunov drift and per-time-slot regret.
With the aid of an auxiliary policy, we bound the drift-plus-regret with a linear function of the total virtual queue backlog size. Finally, by iterated expectations and telescoping sums, we complete the proof.
Details of the proof are given in Appendix \ref{proof: feasibility}.

\textit{Remark 1:}
	Theorem \ref{theorem: feasibility} shows that LAGO is feasible to problem (\ref{p: initial}) when $\boldsymbol{b}$ is interior to the maximal feasibility region. Moreover, the total virtual queue backlog increases linearly as the value of parameter $V$ increases. This implies that the long-term time-averaged node energy consumptions would approach the energy budget vector $\boldsymbol{b}$ as the value of parameter $V$ becomes sufficiently large.

\subsection{Regret Bound}

The following theorem provides an upper bound for the regret $R(T)$ under LAGO.
\begin{theorem}\label{theorem: regret bound}
	Under LAGO, the regret of $T$ time slots defined in (\ref{def: regret}) is upper bounded as follows:
	\begin{align} \label{ineq: regret upper bound}
		R(T)\leq &B/V+(3/(2T)+\sqrt{6a_{\text{max}}\left(N+1\right)(\log T)/T})\theta_{1} \nonumber \\
		&+(3/(2T)+\sqrt{6a_{\text{max}}N(\log T)/T})\theta_{2},
	\end{align}
	where $B\!\triangleq\!\sum_{n\in\mathcal{N}}b_{n}^{2}/2+a_{\text{max}}^{2}(\max\{ \kappa_{\text{max}}^{2}w_{\text{max}}^{2},\eta_{\text{max}}^{2}l_{\text{max}}^{2}\}/2 +N^{2}a_{\text{max}}^{2}\kappa_{\text{max}}^{2}w_{\text{max}}^{2}/2$, $\theta_{1}\!\triangleq\! 2w_{\text{max}}\phi_{\text{max}}a_{\text{max}}$, $\theta_{2}\!\triangleq \!2l_{\text{max}}\rho_{\text{max}}a_{\text{max}}$.
\end{theorem}

We prove Theorem \ref{theorem: regret bound} by following the regret analysis idea in \cite{li2019combinatorial}.
First, we bound the drift-plus-regret developed in the proof of Theorem \ref{theorem: feasibility} with a linear function of the differences between the offloading decisions under LAGO and the optimal policy. 
By adopting iterated expectations and telescoping sums, and introducing another elaborately designed auxiliary policy, we obtain an upper bound for the regret. 
The upper bound is a linear function of the differences between the offloading decisions under LAGO and the auxiliary policy. 
Next, we decompose such a linear bound into different terms. Finally, we adopt techniques such as Chernoff-Hoeffding bound and Jensen's inequality to bound each of them and complete the proof.
More details of the proof are given in Appendix \ref{proof: regret bound}.

\textit{Remark 2:}
	The first term $B/V$ in (\ref{ineq: regret upper bound}) is incurred by the online control procedure, while the second and the third terms are derived from the online learning of processing latency and transmission latency, respectively. 
	As shown by Theorem \ref{theorem: regret bound}, the regret upper bound depends on the time horizon length $T$ and parameter $V$. As $T$ increases to infinity, the regret bound decreases to $B/V$ since the last two terms in the regret bound are of the order $O(\sqrt{(\log T)/T})$. 
	As the value of $V$ increases, the first term $B/V$ decreases and LAGO can achieve a smaller regret.
	When applied in real systems, the selection of the value of $V$ depends on the design tradeoff of the systems.

Note that in our work, we assume that the distribution parameters $\rho_{n}$ and $\phi_{n}$ remain constant over time. 
For the more general case in which environment dynamics are non-stationary with unknown breakpoints, breakpoint detection techniques\cite{ditzler2015learning} can be applied.

\section{Simulation Results}\label{sec: simulation}

\subsection{Simulation Settings}

We conduct extensive simulations in a fog-assisted IoT system with $20$ fog nodes ($N=20$). 
In each time slot, the IoT device has access to a subset of fog nodes with a fixed size $N_{a}=10$. 
During each time slot, we select the subset of accessible fog nodes by sampling $N_{a}$ fog nodes from the fog node set $\mathcal{N}$ uniformly without replacement.
Each simulation is run over $5\times 10^{5}$ time slots ($T=5\times 10^{5}$) based on the commonly adopted settings in fog-assisted IoT systems\cite{you2017energy}\cite{zhang2018femto}, which are specified as follows.

$\diamond$\ \textit{Task arrivals:}
In our simulations, $10$ tasks arrive in each time slot. Therefore, the average task arrival rate is $10$ per time slot. For each new task, its size (in bits) is sampled randomly from the real-world distribution in \cite{mazhar2020characterizing}. The computation intensity of each task is set to be $1000$ CPU cycles per bit.

$\diamond$\ \textit{Transmission rate:}
The transmission rate (bits/s) from the IoT device to fog node in every time slot is sampled from distribution $\mathrm{Unif}(r_{n,\text{min}},r_{n,\text{max}})$, where the values of $r_{n,\text{min}}$ and $r_{n,\text{max}}$ are sampled from distribution $\mathrm{Unif}(5\times 10^{6}, 1.5\times 10^{7})$ and $\mathrm{Unif}(5\times 10^{7}, 1.5\times 10^{8})$, respectively.

$\diamond$\ \textit{Processing rate:}
In every time slot, the task processing rate (cycles/s) on the IoT device is sampled from distribution $\mathrm{Unif}(10^{9},10^{10})$. The task processing rate (cycles/s) on fog node $n$ is sampled from distribution $\mathrm{Unif}(f_{n,\text{min}},f_{n,\text{max}})$, where the values of constants $f_{n,\text{min}}$ and $f_{n,\text{max}}$ are sampled from distribution $\mathrm{Unif}(5\times 10^{9}, 1.5\times 10^{10})$ and $\mathrm{Unif}(1.5\times 10^{10}, 2.5\times 10^{10})$, respectively. 

$\diamond$\ \textit{Energy consumption:}
During every time slot, the unit processing energy (J/cycle) on IoT device is sampled from distribution $\mathrm{Unif}(10^{-10}, 5\times 10^{-10})$. 
The unit processing energy (J/cycle) on each fog node is sampled from distribution $\mathrm{Unif}(5\times 10^{-9}, 1.5\times 10^{-8})$.
The unit transmission energy (J/bit) from IoT device to each fog node $n$ is sampled from distribution $\mathrm{Unif}(10^{-7}, 10^{-6})$.
The energy budget is set as $b_{n}=0.5$ J for each node $n\in\mathcal{N}$.

\subsection{Energy Consumption of Each Node under LAGO}

\begin{figure}[!t]
	\centering
	\includegraphics[scale=0.22]{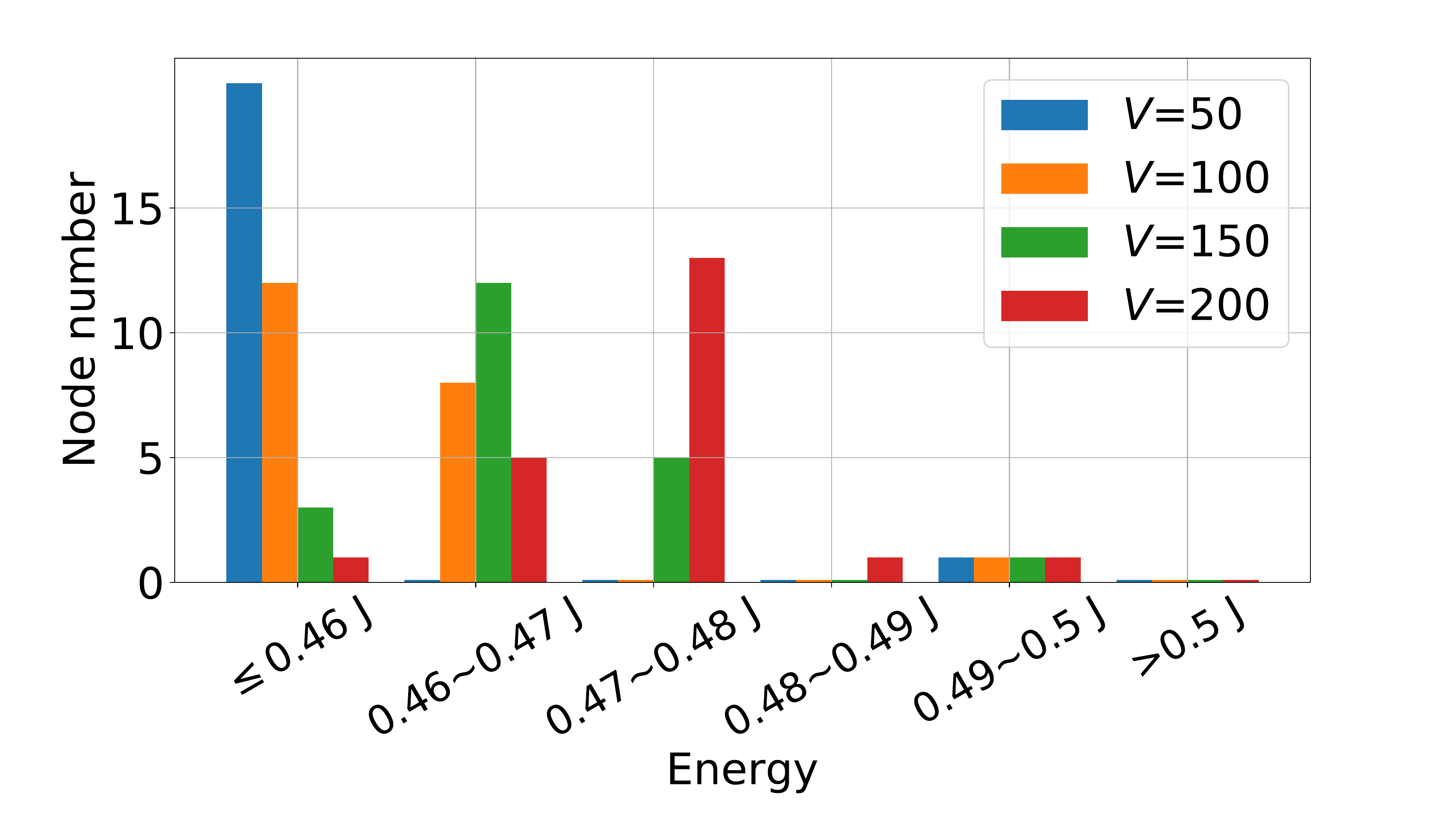}
	\caption{Effect of $V$ on node energy consumption.}
	\label{fig: histogram of node energy} 
\end{figure}

In Figure \ref{fig: histogram of node energy}, we investigate the IoT device's energy consumption under LAGO with different values of $V$.
For example, the leftmost blue bar indicates that there are $20$ nodes whose average energy consumptions are no greater than $0.46$ J when $V=50$.
From the figure we see that when $V=50$, there is only one node that consumes more than $0.47$ J energy. When $V=200$, the energy consumptions of more than half of the nodes are greater than $0.47$ J.
Such results imply that as the value of $V$ increases, the IoT device is more willing to offload tasks, thereby incurring more energy consumptions on each node. Nonetheless, we see that none of the nodes incurs energy consumption beyond the budget ($0.5$ J).

\begin{figure*}[!t]
\centering
  \subfigure[Average task latency.]{
    \includegraphics[scale=0.185]{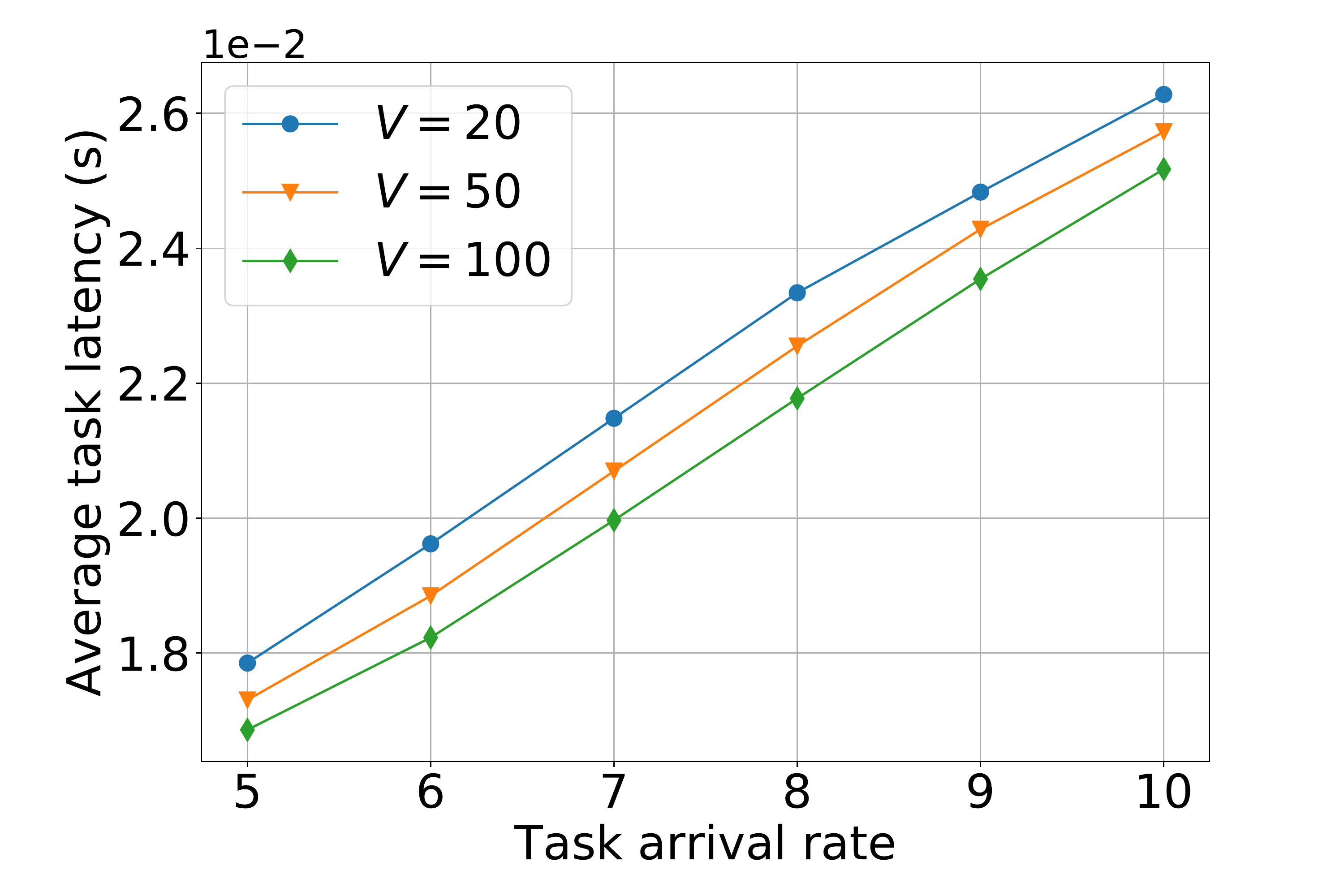} 
  } 
  \subfigure[Total energy consumption.]{ 
	\includegraphics[scale=0.185]{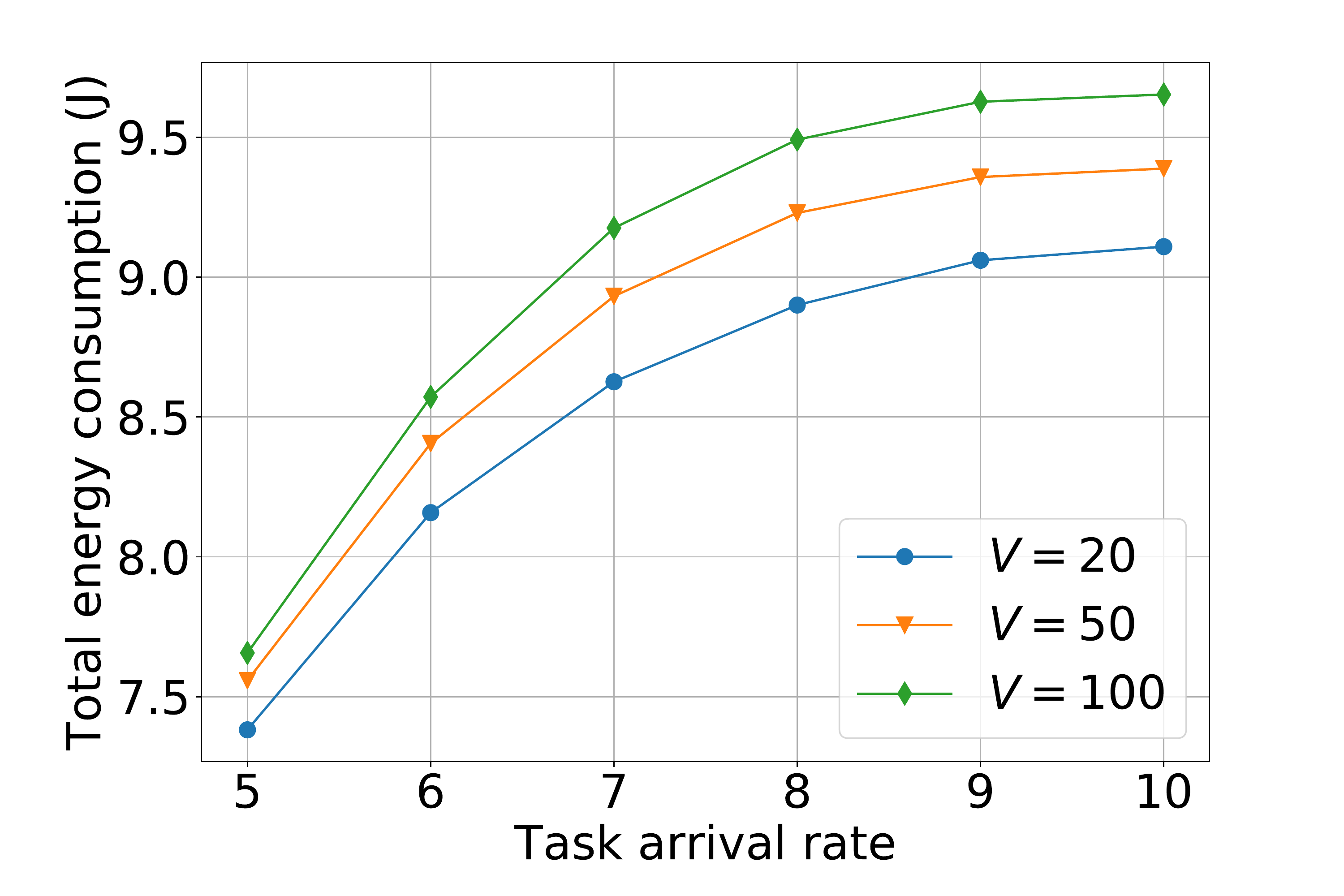}\label{subfig: energy}
  } 
  \subfigure[Regret.]{
    \includegraphics[scale=0.185]{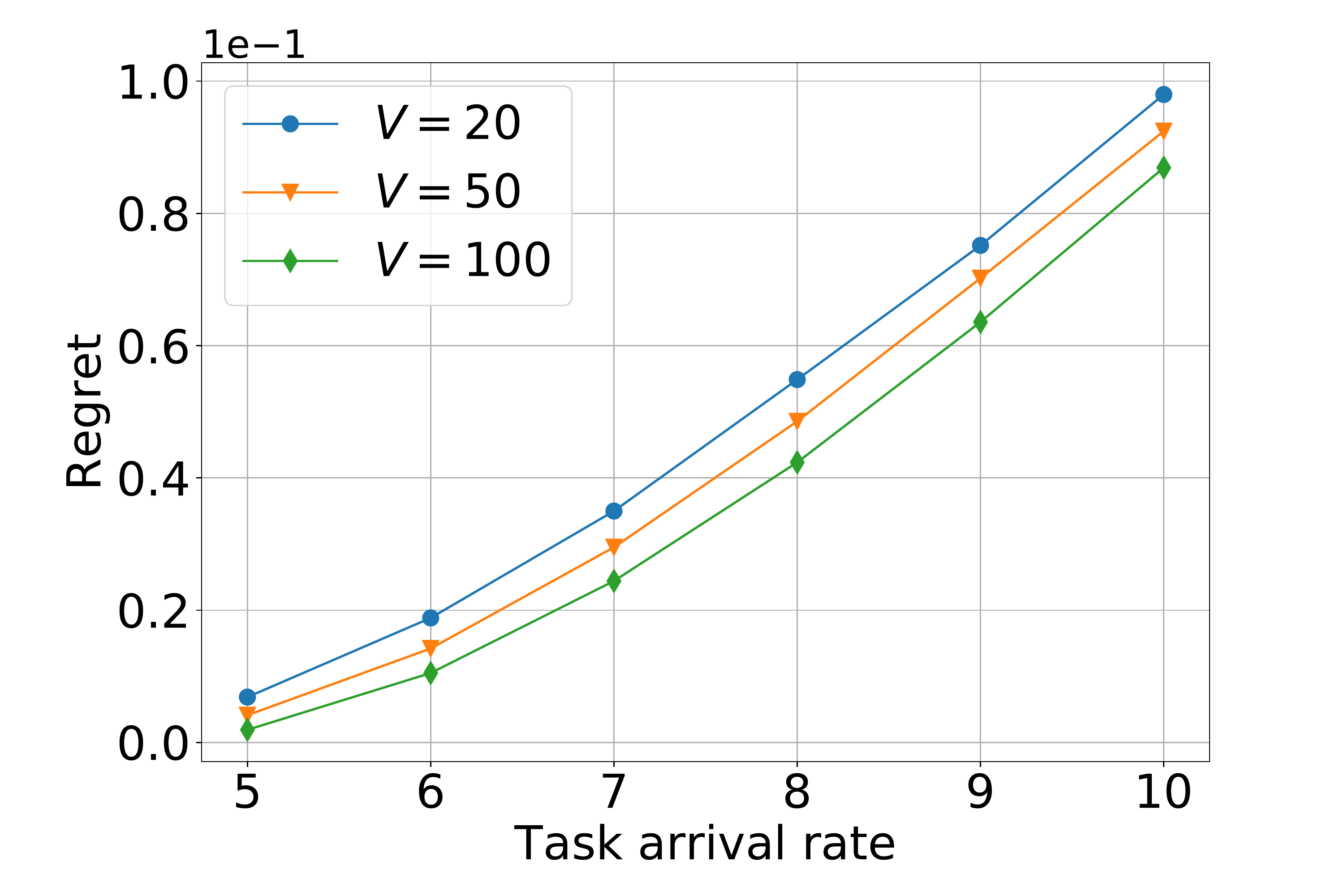} 
  }
  
  \caption{System performance under different task arrival rates.}
  \label{fig: vary arrival}
\end{figure*}

\begin{figure*}[!t]
\centering 
  \subfigure[Average task latency.]{
    \includegraphics[scale=0.185]{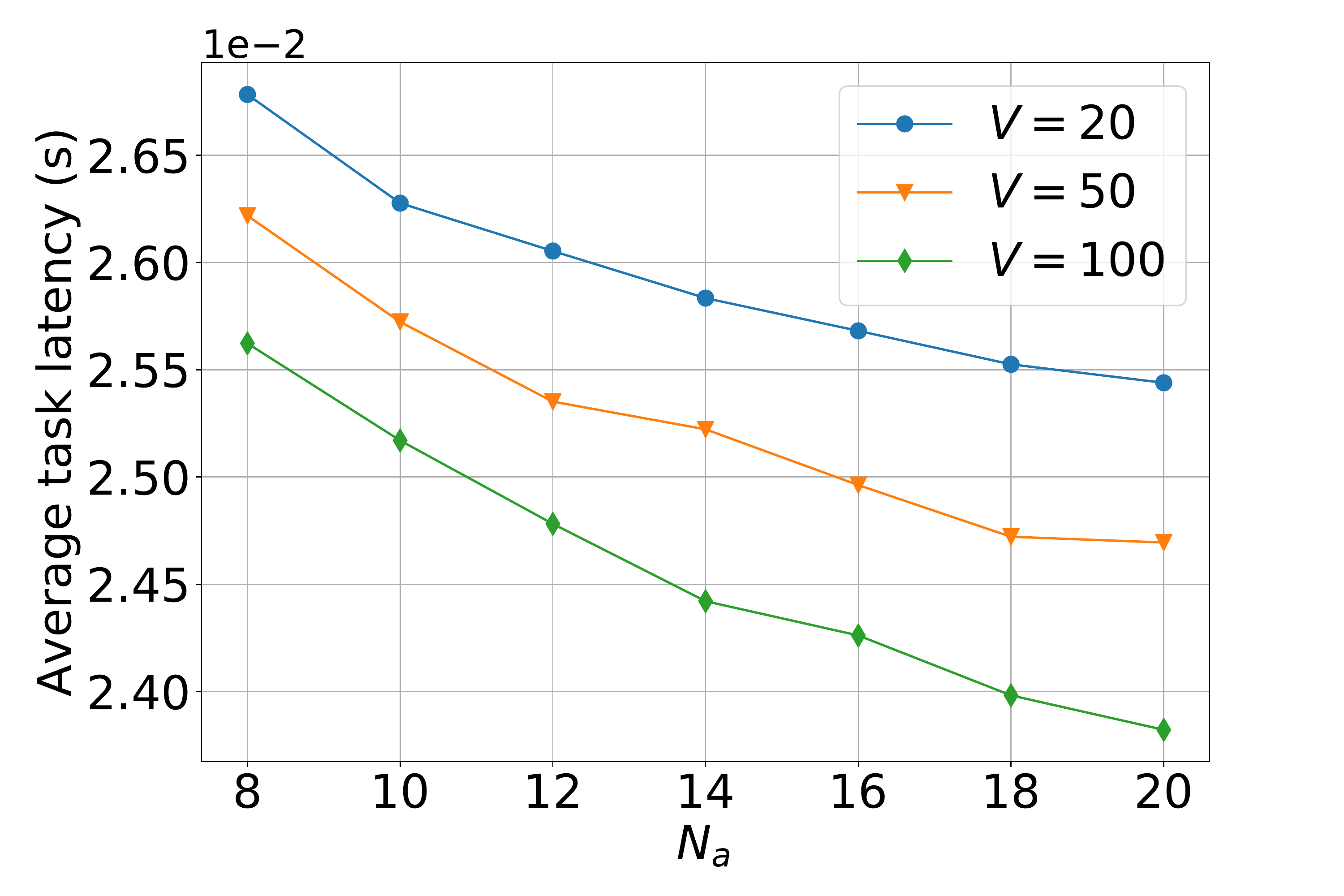} 
  }  
  \subfigure[Total energy consumption.]{
    \includegraphics[scale=0.185]{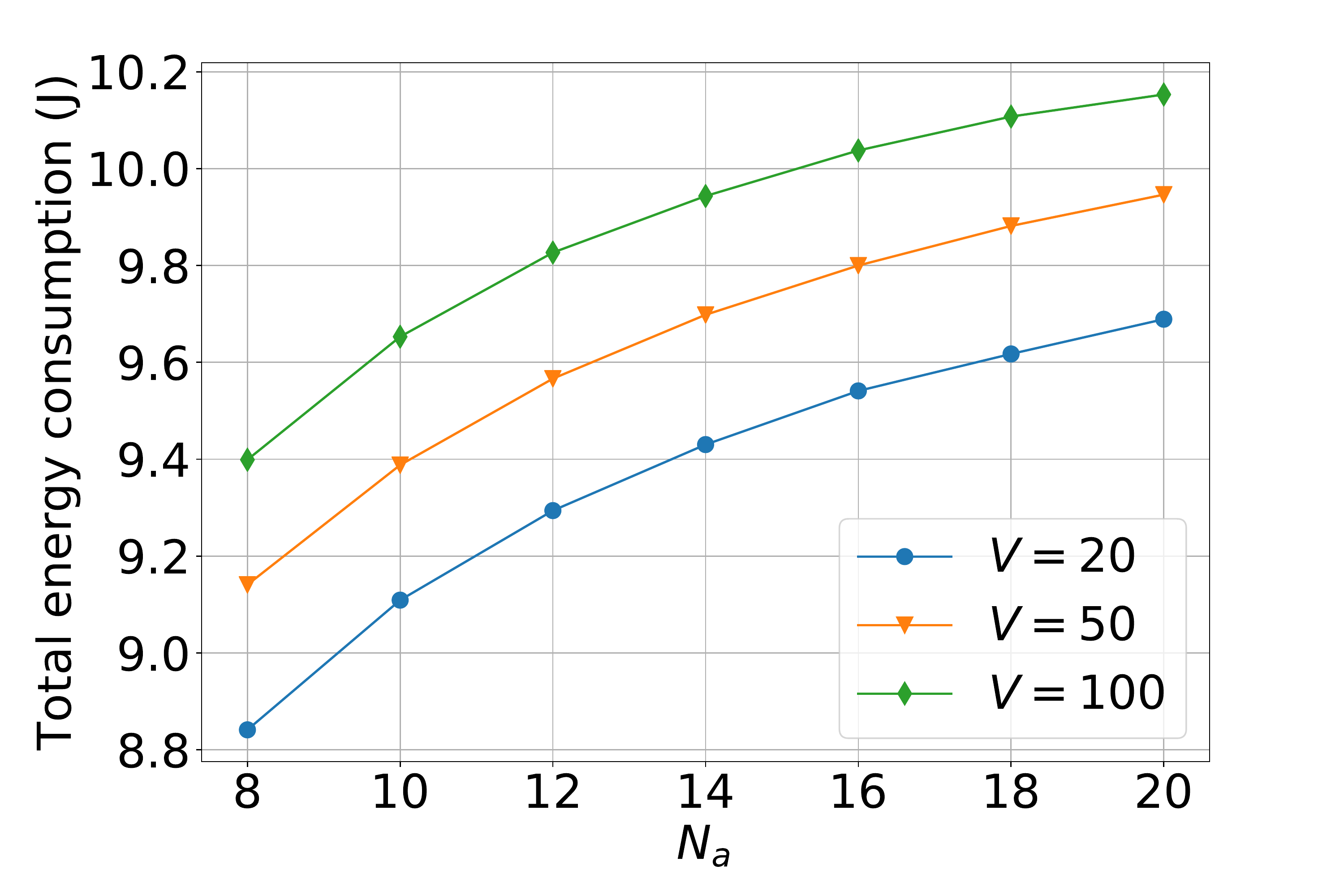} 
  }
  \subfigure[Negative ratio of regret to optimal reward.]{
    \includegraphics[scale=0.185]{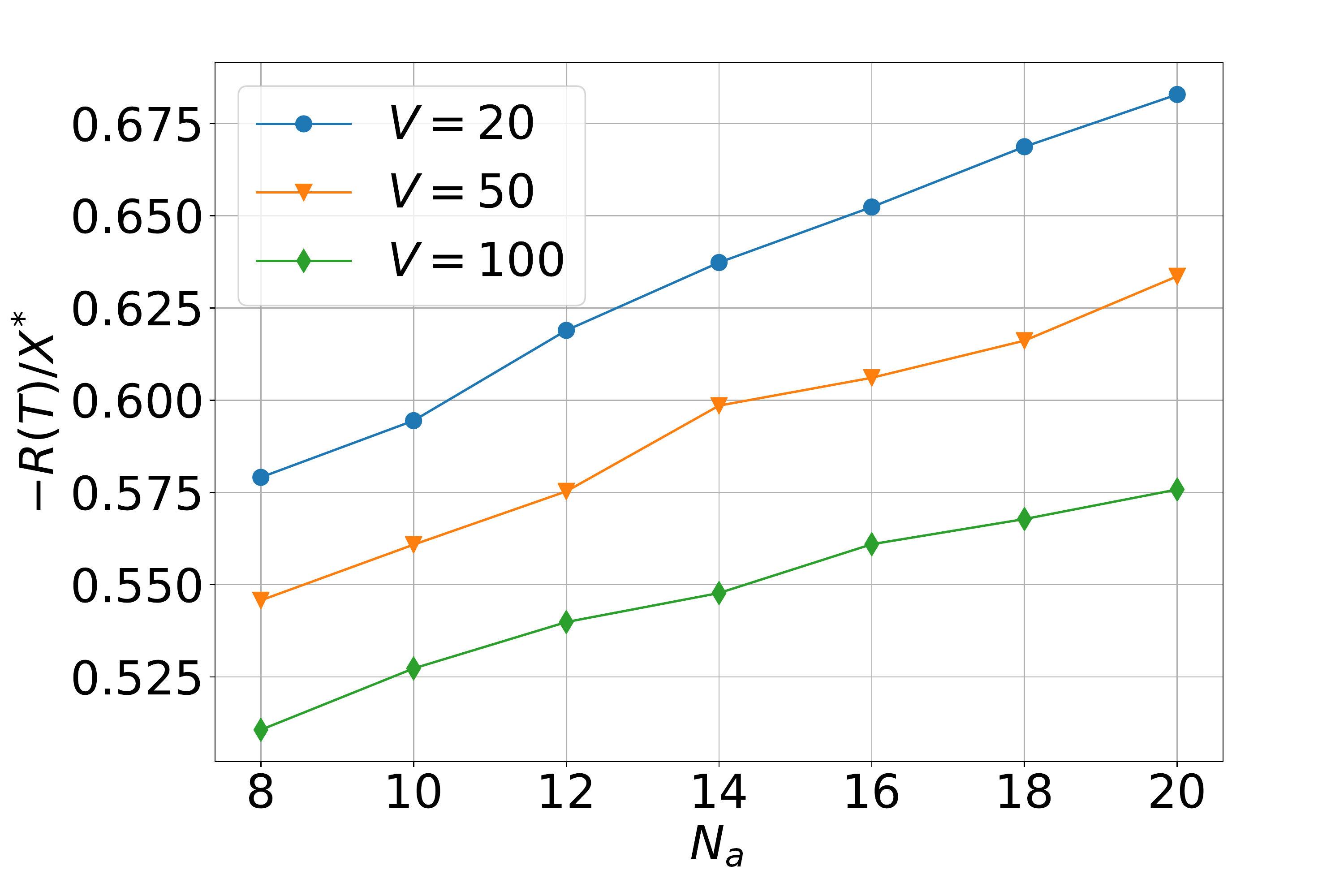} 
  }
  \caption{System performance under different accessible fog subset sizes $N_{a}$.}
  \label{fig: vary fog subset size}
\end{figure*} 

\begin{figure*}[!t]
\centering 
  \subfigure[Total energy consumption.]{ 
	\includegraphics[scale=0.185]{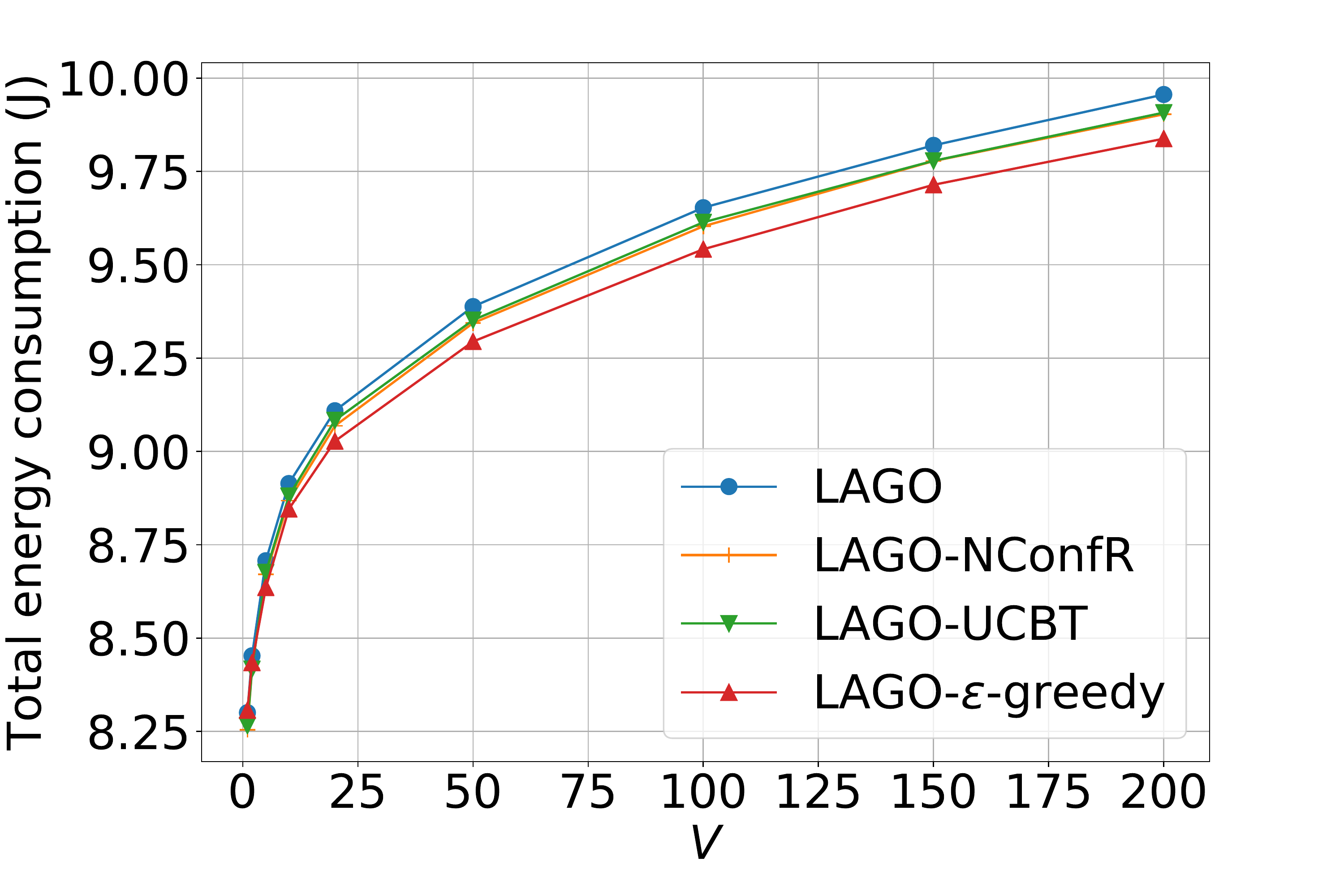}\label{subfig: energy}
  }  
  \subfigure[Regret.]{
    \includegraphics[scale=0.185]{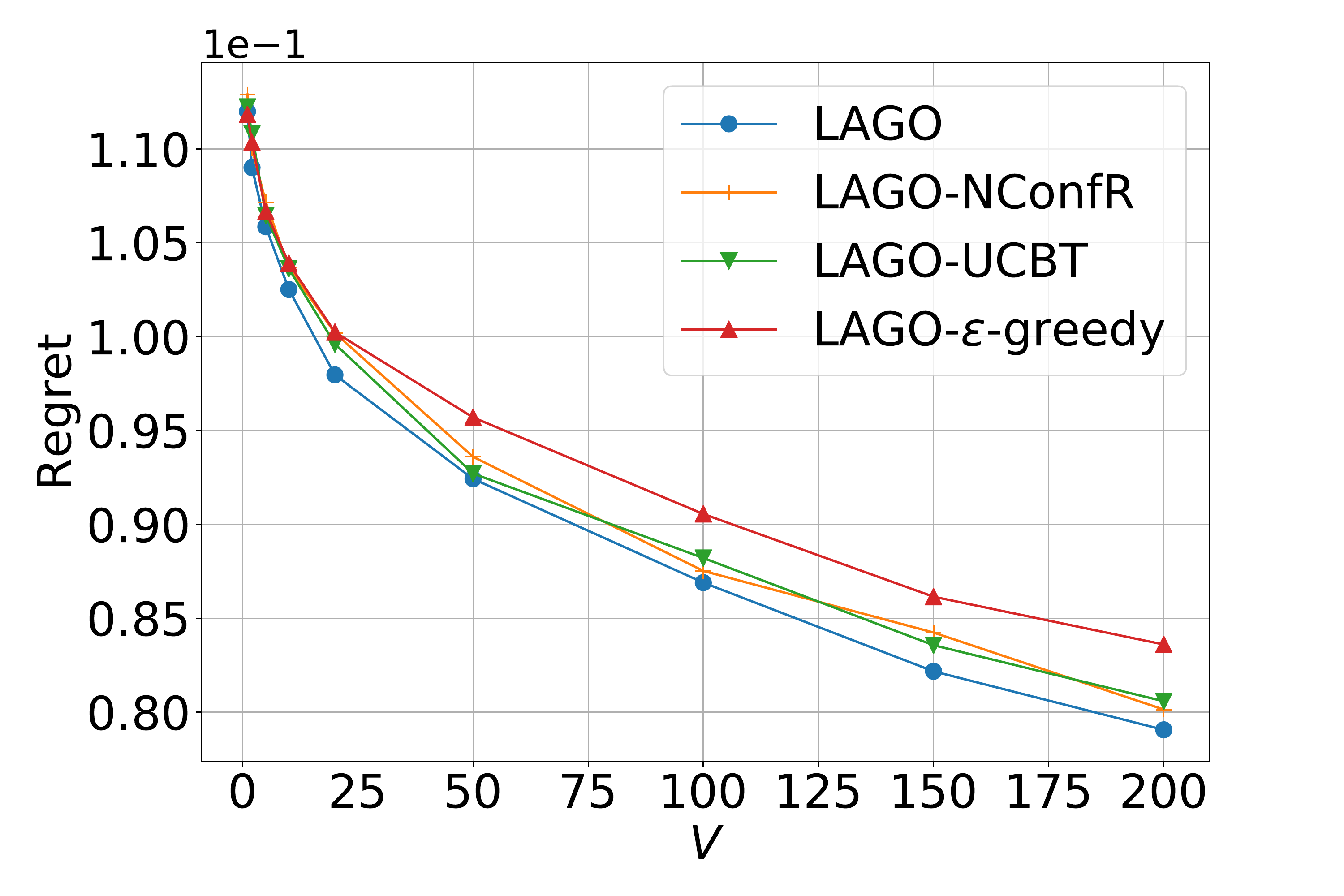} 
  }
  \subfigure[Latency-energy tradeoff.]{
    \includegraphics[scale=0.185]{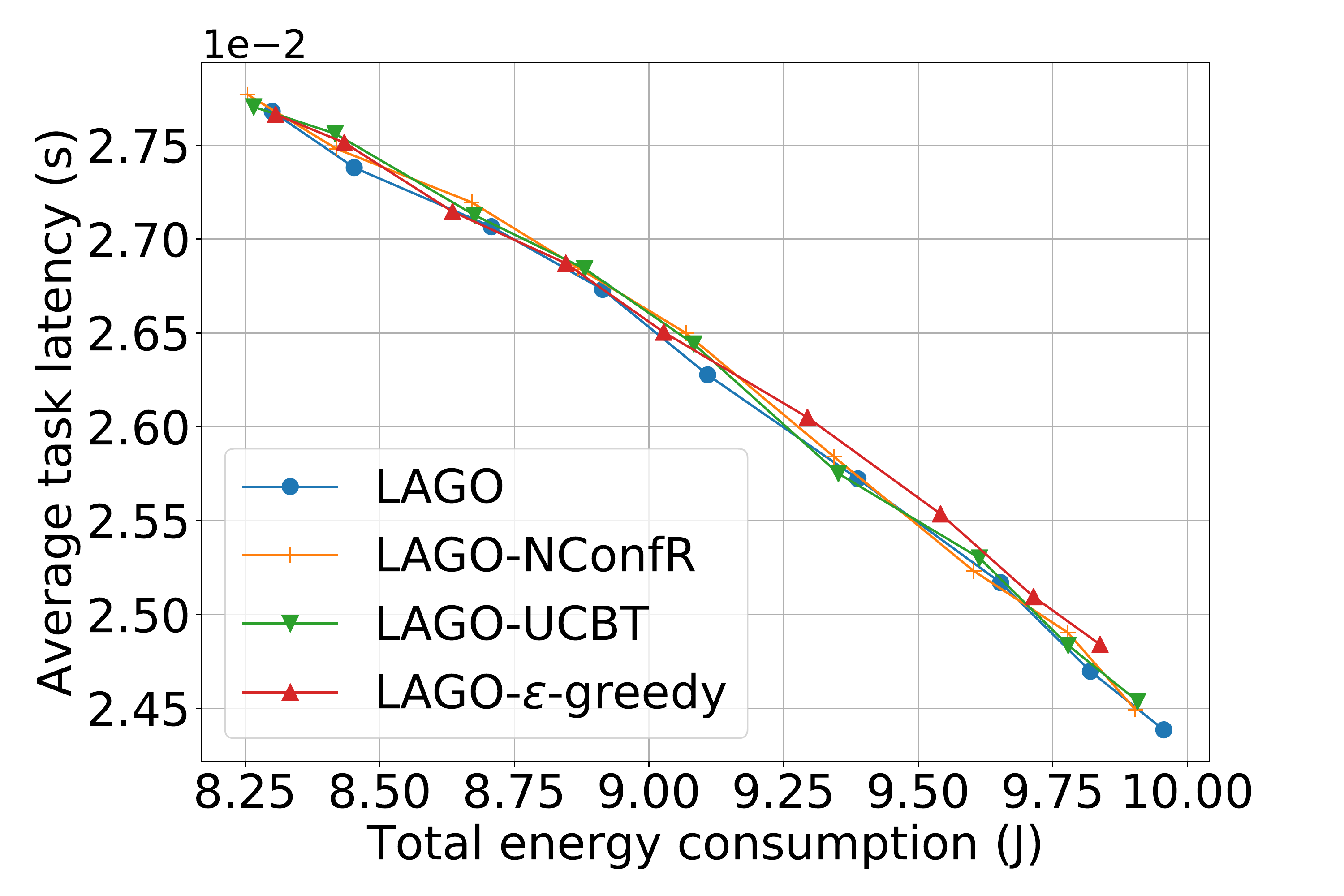} 
  }
  
  \caption{Comparison of the performance of LAGO and its variants.}
  \label{fig: LAGO performance}
\end{figure*}

\subsection{System Performance under Different Task Arrival Rates}

In this section, we investigate the effects of the task arrival rate on the system performance under LAGO. 
By varying the values of $a_{\text{max}}$, we obtain the simulation result in Figure \ref{fig: vary arrival}. 

Under a fixed value of $V$, we have the following observations.
Figure \ref{fig: vary arrival}(a) shows that when the task arrival rate increases, the average latency of each task grows. The result is because there are more tasks competing for the limited system resources under a larger task arrival rate.
Figure \ref{fig: vary arrival}(b) implies that the total energy consumption grows with the increase of the task arrival rate since there are more tasks to be served. 
However, the growth rate gradually goes down with the increase of the task arrival rate. For example, when $V=100$, the total energy consumption increases by $19.84\%$ as the task arrival rate increases from $5$ to $7$, but it increases by only $1.70\%$ as the task arrival rate increases from $8$ to $10$.
This is because the service capacity of the system is limited and the system can not serve more tasks when the resources are exhausted.
Figure \ref{fig: vary arrival}(c) verifies our theoretical analysis in (\ref{ineq: regret upper bound}) that the regret increases as the task arrival rate increases.

\subsection{Effect of Parameter $V$}
Figure \ref{fig: vary arrival} also illustrates the impact of the value of $V$ on the system performance. The figure shows that under the same task arrival rate, when the value of $V$ increases, the average task latency decreases while the total energy consumption increases. 
The reason is that under LAGO, with a larger value of $V$, the IoT device prefers shortening task latency to reducing energy consumption. As a result, when making an offloading decision, the IoT device will select the low-latency candidate regardless of the potential consequent high energy consumption. For example, offloading tasks to fog nodes usually reduces task latencies while consuming extra wireless transmission energy.
The results also verify our remarks in Section \ref{sec: algorithm}.
In practice, one can select the value of $V$ to adjust the tradeoff between shortening task latencies and reducing energy consumptions based on its system design objective.
As for regret, it has the same changing trend as the average task latency. This is because they have a positive linear relationship under the fixed task arrival rate according to (\ref{def: regret}).

\subsection{System Performance under Different Accessible Fog Node Numbers}

In this section, we investigate the system performance under LAGO with respect to $N_{a}$ (the size of the subset of accessible fog nodes in each time slot) from $8$ to $20$ under different values of $V$. 
 According to Figure \ref{fig: vary fog subset size}(a), the average task latency reduces as the value of $N_{a}$ grows. For example, under the setting of $V=100$, the average task latency decreases by $7.03\%$ as $N_{a}$ increases from $8$ to $20$.
The reason is that with a larger accessible fog node subset, the IoT device is able to select fog nodes with shorter transmission and processing latencies.
However, from Figure \ref{fig: vary fog subset size}(b), we observe an opposite trend in the change of the time-averaged total energy consumption of all nodes as the value of $N_{a}$ grows. The results imply the tradeoff between task latency and energy consumption. 
Such a tradeoff exists because the minimal-latency choice for the offloading decision is usually not the one that consumes the least energy.

Figure \ref{fig: vary fog subset size}(c) presents the negative ratio of regret $R(T)$ to the optimal expected time-averaged reward $X^{*}$ with different $N_{a}$. Recall that regret $R(T)$ is defined as the gap between the optimal reward $X^{*}$ and the achieved expected time-averaged reward under LAGO, as shown in (\ref{def: regret}). In other words, it represents the reward loss incurred by making suboptimal offloading decisions under uncertainties with respect to $X^{*}$. To compare the relative reward loss (\textit{i.e.}, \textit{relative regret}) under different values of $N_{a}$, we consider the negative regret-optimal reward ratio $-R(T)/X^{*}$ instead of regret $R(T)$.
The negative symbol here is to ensure the positiveness of the ratio.\footnote{The value of reward is negative by the definition of reward in (\ref{def: task reward}).} The figure shows that the negative ratio grows with the increase of $N_{a}$. 
Intuitively, when there are more fog node candidates, LAGO has a higher probability to select the suboptimal one, resulting in a larger relative regret.

\subsection{LAGO vs. its Variants}

Recall that we can substitute UCB1 with other bandit learning methods in LAGO. In our simulations, we consider the following variants of LAGO:
LAGO-UCBT,
LAGO-$\epsilon$-greedy,
and LAGO-NConfR.
\begin{itemize}
	\item \textbf{LAGO-UCBT}: LAGO-UCBT replaces UCB1 estimate with UCB-tuned\cite{auer2002finite} estimate in the online learning phase of LAGO (lines \ref{algline: estimate 1} and \ref{algline: estimate 2} in Algorithm \ref{alg: policy}) while the rest of the algorithm remains the same as LAGO.
	\item \textbf{LAGO-$\epsilon$-greedy}: LAGO-$\epsilon$-greedy differs from LAGO in the node selection phase (line \ref{algline: node selection} in Algorithm \ref{alg: policy}). Under LAGO-$\epsilon$-greedy, for each task, with a probability of $\epsilon=0.1$, the IoT device uniformly randomly selects a node from the set of available nodes. Otherwise, it selects the empirically best node, \textit{i.e.}, the node with the minimal value function $v_{i,n}(t,\bar{\rho}_{n}(t),\bar{\phi}_{n}(t))$\cite{vermorel2005multi}.
	\item \textbf{LAGO-NConfR}: Different from LAGO in the node selection phase (line \ref{algline: node selection} in Algorithm \ref{alg: policy}), LAGO-NConfR sets the confidence radius in UCB1 estimates (\ref{eq: UCB estimate about transmit}) and (\ref{eq: UCB estimate about process}) to be zero. More specifically, under LAGO-NConfR, the IoT device always selects the node with the minimal value function $v_{i,n}(t,\bar{\rho}_{n}(t),\bar{\phi}_{n}(t))$ from set $\mathcal{N}(t)$ for each task $i\in\mathcal{A}(t)$. In our simulations, we call such a scheme LAGO-NConfR (No Confidence Radius). Intuitively, under such a scheme, the exploration procedure is dropped from online learning.
\end{itemize}

The performance of LAGO and its variants is illustrated in Figure \ref{fig: LAGO performance}.
Specifically, Figures \ref{fig: LAGO performance}(a) and \ref{fig: LAGO performance}(b) show that when the value of $V$ changes, all schemes have the same changing trends in the total energy consumptions and regret. Under the same value of $V$, LAGO achieves the minimal regret but the most total energy consumptions. However, as shown in Figure \ref{fig: LAGO performance}(c), all schemes achieve similar latency-energy tradeoffs.


An interesting result is that even without exploration in the learning part, LAGO-NConfR performs comparably against LAGO and other variants. The reason is that under LAGO-NConfR, each node will not be sampled too frequently to ensure the stability of its corresponding virtual queue. 
In other words, even without exploration in the learning part, LAGO-NConfR will explore other nodes when the virtual queue of the empirically best node is too large, \textit{i.e.}, the control part of LAGO-NConfR still enforces exploration implicitly.

\section{Conclusion}\label{sec: conclusion}

In this paper, we proposed LAGO, an online task offloading scheme, to address the problem of task offloading in fog-assisted IoT systems with unknown node processing capacities. 
Our theoretical results showed that LAGO achieves a tunable latency-energy tradeoff under the energy constraints with an $O(1/V+\sqrt{(\log T)/T})$ regret bound over time horizon $T$. By conducting extensive simulations, we showed the performance of LAGO, which verified our theoretical results.

\ifCLASSOPTIONcaptionsoff
  \newpage
\fi

\bibliographystyle{IEEEtran}
\bibliography{references.bib, IEEEabrv}

\appendices

\section{Proof of Theorem \ref{theorem: feasibility}}\label{proof: feasibility}

First, we define a Lyapunov function $L\left(\boldsymbol{Q}\left(t\right)\right)$ as
\begin{equation}\label{def: Lyapunov func}
	L\left(\boldsymbol{Q}\left(t\right)\right)\triangleq \frac{1}{2}\sum_{n\in\mathcal{N}}\left(Q_{n}\left(t\right)\right)^{2},
\end{equation}
in which $\boldsymbol{Q}(t)=(Q_{0}(t),Q_{1}(t),\dots,Q_{N}(t))$ is the vector of all virtual queue backlogs in time slot $t$.
Then we define the corresponding Lyapunov drift $\Delta L\left(\boldsymbol{Q}\left(t\right)\right)$ as
\begin{equation} \label{def: Lyapunov drift}
	\Delta L\left(\boldsymbol{Q}\left(t\right)\right)\triangleq\mathbb{E}\left[L\left(\boldsymbol{Q} \left(t+1\right)\right)-L\left(\boldsymbol{Q}\left(t\right)\right)\vert\boldsymbol{Q}\left(t\right)\right].
\end{equation}
To develop an upper bound for $\Delta L\left(\boldsymbol{Q}\left(t\right)\right)$, we consider the difference of $L(\boldsymbol{Q}(t))$ between successive time slots:
\begin{equation}\label{ineq: difference bound 1}
\begin{split}
	&L\left(\boldsymbol{Q}\left(t+1\right)\right)-L\left(\boldsymbol{Q}\left(t\right)\right)\\
	&\leq \frac{1}{2}\sum_{n\in\mathcal{N}}[(Q_{n}(t)-b_{n})^{2}+(E_{n}(t))^{2}+2Q_{n}(t)E_{n}(t)]\\
	&~~~-\frac{1}{2}\sum_{n\in\mathcal{N}}(Q_{n}(t))^{2}\\
	&=\frac{1}{2}\sum_{n\in\mathcal{N}}(b_{n}^{2}+(E_{n}(t))^{2})+\sum_{n\in\mathcal{N}}Q_{n}(t)(E_{n}(t)-b_{n}).
\end{split}
\end{equation}
According to (\ref{eq: user energy}), (\ref{eq: fog energy}) and the boundedness assumptions of the model, we have $E_{0}\left(t\right)\leq a_{\text{max}}\max\left\{ \kappa_{\text{max}}w_{\text{max}},\eta_{\text{max}}l_{\text{max}}\right\} $ and $E_{n}\left(t\right)\leq a_{\text{max}}\kappa_{\text{max}}w_{\text{max}}$, thus
\begin{equation}
\begin{split}
	&\frac{1}{2}\sum_{n\in\mathcal{N}}\left(b_{n}^{2}+\left(E_{n}\left(t\right)\right)^{2}\right)\\
	&\leq \frac{1}{2}a_{\text{max}}^{2}\max\left\{ \kappa_{\text{max}}^{2}w_{\text{max}}^{2},\eta_{\text{max}}^{2}l_{\text{max}}^{2}\right\} \\
	&~~~+\frac{1}{2}N^{2}a_{\text{max}}^{2}\kappa_{\text{max}}^{2}w_{\text{max}}^{2}+\frac{1}{2}\sum_{n\in\mathcal{N}}b_{n}^{2} \triangleq B.
\end{split}
\end{equation}
It follows by (\ref{ineq: difference bound 1}) that 
\begin{equation}\label{ineq: difference bound 2}
	L(\boldsymbol{Q}(t+1))-L(\boldsymbol{Q}(t))\leq B+\sum_{n\in\mathcal{N}}Q_{n}(t)(E_{n}(t)-b_{n}).
\end{equation}
Then we obtain the upper bound of Lyapunov drift as
\begin{equation}\label{ineq: drift bound}
	\Delta L(\boldsymbol{Q}(t))\leq B+\mathbb{E}\left[\sum_{n\in\mathcal{N}}Q_{n}(t)(E_{n}(t)-b_{n})|\boldsymbol{Q}(t)\right].
\end{equation}
Now suppose the optimal policy makes offloading decision $\{I_{i}^{*}(t)\}_{i\in\mathcal{A}(t)}$ in each time slot $t$, then the optimal expected reward is
\begin{equation}\label{eq: optimal reward}
	X^{*}=\frac{1}{T}\sum_{t=0}^{T-1}\sum_{i\in\mathcal{A}(t)}\mathbb{E}[X_{i,I_{i}^{*}(t)}(t)].
\end{equation}
We can rewrite the regret $R(T)$ under policy $\{I_{i}(t)\}_{i,t}$ as
\begin{equation}\label{eq: regret 1}
\begin{split}
	R(T)=\frac{1}{T}\sum_{t=0}^{T-1}\sum_{i\in\mathcal{A}(t)}(\mathbb{E}[X_{i,I_{i}^{*}(t)}(t)]-\mathbb{E}[X_{i,I_{i}(t)}(t)]).
\end{split}
\end{equation}
Since $X_{i,n}(t)=-D_{i,n}(t)$, it follows that
\begin{equation}
	R(T)=\frac{1}{T}\sum_{t=0}^{T-1}\sum_{i\in\mathcal{A}(t)}(\mathbb{E}[D_{i,I_{i}(t)}(t)]-\mathbb{E}[D_{i,I_{i}^{*}(t)}(t)]).
\end{equation}
We define the one-slot regret as 
\begin{equation}\label{def: one-slot regret}
	\Delta R\left(t\right)\triangleq\sum_{i\in\mathcal{A}(t)}\left(\mathbb{E}\left[D_{i,I_{i}\left(t\right)}\left(t\right)\right]-\mathbb{E}\left[D_{i,I_{i}^{*}\left(t\right)}\left(t\right)\right]\right).
\end{equation}
Then the regret $R(T)$ can be expressed as
\begin{equation}\label{eq: regret 2}
	R\left(T\right)=\frac{1}{T}\sum_{t=0}^{T-1}\mathbb{E}\left[\Delta R\left(t\right)\right].
\end{equation}
Note that $\Delta R(t)\geq 0$ for all $t$ since the optimal policy always choose the node which can achieve the minimal expected task latency. 
Substitute (\ref{eq: task latency}) into (\ref{def: one-slot regret}), we have
\begin{equation}
\begin{split}
	\Delta R\left(t\right)=	&\!\sum_{i\in\mathcal{A}(t)}\!\Bigg(\mathbb{E}\left[\frac{L_{i}\left(t\right)\!\mathds{1}\!\left\{ I_{i}\left(t\right)>0\right\} }{R_{i,I_{i}\left(t\right)}\left(t\right)}\!+\!\frac{W_{i}\left(t\right)}{F_{i,I_{i}\left(t\right)}\left(t\right)}\right]\\
	&-\mathbb{E}\left[\frac{L_{i}\left(t\right)\!\mathds{1}\!\left\{ I_{i}^{*}\left(t\right)>0\right\} }{R_{i,I_{i}^{*}\left(t\right)}\left(t\right)}+\frac{W_{i}\left(t\right)}{F_{i,I_{i}^{*}\left(t\right)}\left(t\right)}\right]\!\Bigg).
\end{split}
\end{equation}
Since $\mathbb{E}[1/R_{i,n}(t)]=\rho_{n}$ and $\mathbb{E}[1/F_{i,n}(t)]=\phi_{n}$, given offloading decision $\{I_{i}(t)\}_{i}$, we have
\begin{equation}\label{def: one-slot regret}
\begin{split}
	&\Delta R(t)\!=\! \sum_{i\in\mathcal{A}(t)}(\rho_{I_{i}(t)}L_{i}(t)\mathds{1}\{ I_{i}(t)>0\} +\phi_{I_{i}(t)}W_{i}(t))\\
	&~~~~~-\sum_{i\in\mathcal{A}(t)}(\rho_{I_{i}^{*}(t)}L_{i}(t)\mathds{1}\{ I_{i}^{*}(t)>0\} +\phi_{I_{i}^{*}(t)}W_{i}(t)).
\end{split}
\end{equation}
For simplicity of expression, we let 
\begin{equation}\label{def: D*}
D^{*}(t)\triangleq\sum_{i\in\mathcal{A}(t)}\rho_{I_{i}^{*}(t)}L_{i}(t)\mathds{1}\{ I_{i}^{*}(t)>0\} +\phi_{I_{i}^{*}(t)}W_{i}(t),
\end{equation}
then we have
\begin{equation}\label{def: simplified one-slot regret}
\begin{split}
	\Delta R(t)=& \sum_{i\in\mathcal{A}(t)}\!\left(\rho_{I_{i}(t)}L_{i}(t)\mathds{1}\{ I_{i}(t)>0\}\!+\!\phi_{I_{i}(t)}W_{i}(t)\right)\\
	&-D^{*}(t).
\end{split}
\end{equation}
Next, define the drift-plus-regret $\Delta_{V}\left(\boldsymbol{Q}\left(t\right)\right)$ as 
\begin{equation}\label{def: drift-plus-penalty}
	\Delta_{V}\left(\boldsymbol{Q}\left(t\right)\right)\triangleq \Delta\left(\boldsymbol{Q}\left(t\right)\right)+V\mathbb{E}\left[\Delta R\left(t\right)|\boldsymbol{Q}\left(t\right)\right],
\end{equation}
which is the weighted sum of Lyapunov drift and expected one-slot regret. 
According to (\ref{ineq: drift bound}), $\Delta_{V}\left(\boldsymbol{Q}\left(t\right)\right)$ can be upper bounded as follows:
\begin{multline}\label{ineq: general drift-plus-regret bound}
	\Delta_{V}\left(\boldsymbol{Q}\left(t\right)\right)\leq B\\
	+\mathbb{E}\left[\sum_{n\in\mathcal{N}}Q_{n}\left(t\right)\left(E_{n}\left(t\right)-b_{n}\right)+V\Delta R\left(t\right)\Bigg\vert\boldsymbol{Q}\left(t\right)\right].
\end{multline}
We can rewrite the right-hand side of (\ref{ineq: general drift-plus-regret bound}) by (\ref{eq: user energy}), (\ref{eq: fog energy}), and (\ref{def: simplified one-slot regret}) as follows:
\begin{equation}\label{form: detailed drift-plus-regret bound}
\begin{split}
	&\Delta_{V}\left(\boldsymbol{Q}\left(t\right)\right)\leq B-V\mathbb{E}\left[D^{*}\left(t\right)|\boldsymbol{Q}\left(t\right)\right]-\sum_{n\in\mathcal{N}}b_{n}Q_{n}\left(t\right)\\
	&+\mathbb{E}\Bigg[\sum_{i\in\mathcal{A}\left(t\right)}\left(Q_{0}\left(t\right)\kappa_{0}\left(t\right)W_{i}\left(t\right)+V\phi_{0}W_{i}\left(t\right)\right)\\
	&~~~~~~~~~~~~~~~~~~~~~~~~~~~~~~~~~~~~\cdot\mathds{1}\left\{ I_{i}\left(t\right)=0\right\} \big\vert\boldsymbol{Q}\left(t\right)\big]\\
	&+\mathbb{E}\Bigg[\!\sum_{i\in\mathcal{A}\left(t\right)}\sum_{n=1}^{N}\big(Q_{n}(t)\kappa_{n}(t)W_{i}(t)+Q_{0}(t)\eta_{n}(t)L_{i}(t)\\
	&~~~~~~~+V(\phi_{n}W_{i}(t)+\rho_{n}L_{i}(t))\big)\mathds{1}\left\{ I_{i}(t)=n\right\} \big\vert\boldsymbol{Q}(t)\big]\\
	&=B-V\mathbb{E}\left[D^{*}\left(t\right)|\boldsymbol{Q}\left(t\right)\right]-\sum_{n\in\mathcal{N}}b_{n}Q_{n}\left(t\right)\\
	&+\mathbb{E}\Bigg[\sum_{i\in\mathcal{A}\left(t\right)}\sum_{n\in\mathcal{N}}v_{i,n}\left(t,\rho_{n},\phi_{n}\right)\mathds{1} \left\{ I_{i}(t)=n\right\}\bigg\vert\boldsymbol{Q}\left(t\right) \Bigg]\\
	&=B-V\mathbb{E}\left[D^{*}\left(t\right)|\boldsymbol{Q}\left(t\right)\right]-\sum_{n\in\mathcal{N}}b_{n}Q_{n}\left(t\right)\\
	&+\!\mathbb{E}\Bigg[\!\sum_{i\in\mathcal{A}(t)}\!\sum_{n\in\mathcal{N}}\!v_{i,n}(t,\hat{\rho}_{n}(t),\hat{\phi}_{n}(t))\mathds{1}\{ I_{i}(t)\!=\!n\}\bigg\vert \boldsymbol{Q}(t)\!\Bigg]\\
	&+V\mathbb{E}\Bigg[\sum_{i\in\mathcal{A}\left(t\right)}\big((\phi_{I_{i}(t)}-\hat{\phi}_{I_{i}(t)}(t))W_{i}(t)\\
	&~~~~~+(\rho_{I_{i}(t)}-\hat{\rho}_{I_{i}(t)}(t))L_{i}(t)\mathds{1}\{I_{i}(t)>0\} \big)\big\vert\boldsymbol{Q}\left(t\right)\big],
\end{split}
\end{equation}
where function $v_{i,n}\left(\cdot\right)$ is defined in (\ref{def: auxiliary func}). The empirical means $\hat{\rho}_{n}(t)$ and $\hat{\phi}_{n}(t)$ are defined in (\ref{eq: UCB estimate about transmit}) and (\ref{eq: UCB estimate about process}), respectively.
Since $\phi_{n}\leq \phi_{\text{max}}$, $L_{i}(t)\leq l_{\text{max}}$, $\rho_{n}\leq\rho_{\text{max}}$, $W_{i}(t)\leq w_{\text{max}}$, and $A(t)\leq a_{\text{max}}$, it follows that
\begin{equation}\label{ineq: drift-plus-regret bound 1}
\begin{split}
	&\Delta_{V}\left(\boldsymbol{Q}\left(t\right)\right)\leq B-V\mathbb{E}\left[D^{*}\left(t\right)|\boldsymbol{Q}\left(t\right)\right]-\sum_{n\in\mathcal{N}}b_{n}Q_{n}\left(t\right)\\
	&+\!\mathbb{E}\Bigg[\!\sum_{i\in\mathcal{A}(t)}\!\sum_{n\in\mathcal{N}}\!v_{i,n}(t,\hat{\rho}_{n}(t),\hat{\phi}_{n}(t))\mathds{1}\{ I_{i}(t)\!=\!n\}\bigg\vert \boldsymbol{Q}(t)\!\Bigg]\\
	&+Va_{\text{max}}\left(\phi_{\text{max}}w_{\text{max}}+\rho_{\text{max}}l_{\text{max}}\right).
\end{split}
\end{equation}

Recall that \textit{$S$-only} policies make \textit{i.i.d.} offloading decisions over time and the decisions in each time slot $t$ only depend on the observable system state $(\eta_{n}(t),\kappa_{n}(t), L_{i}(t), W_{i}(t))_{n,i}$.
Based on the assumption that $\boldsymbol{b}$ is an interior point of the maximal feasibility region, there must exist some constant $\epsilon > 0$ such that $\boldsymbol{b} -\epsilon \boldsymbol{1}$ is also an interior point of the maximal feasibility region. Therefore, there exists an $S$-only policy which makes offloading decision $\{I^{\epsilon}_{i}(t)\}_{i\in\mathcal{A}(t)}$ in each time slot $t$ such that 
\begin{equation}\label{ineq: energy constraint under A-only}
	\mathbb{E}\left[E_{n}^{\epsilon}\left(t\right)\right]+\epsilon\leq b_{n},\ \forall n\in\mathcal{N}
\end{equation}
holds for all time slots, where $E_{n}^{\epsilon}\left(t\right)$ is the energy consumption of node $n$ under offloading decision $\{I_{i}^{\epsilon}(t)\}_{i\in\mathcal{A}(t)}$.
Under our policy LAGO which makes offloading decision $\{I_{i}(t)\}_{i\in\mathcal{A}(t)}$ in each time slot $t$ as shown in (\ref{eq: our policy}), since $I_{i}^{\epsilon}(t)$ is independent of $\boldsymbol{Q}(t)$, we have
\begin{equation}
\begin{split}
	&\Delta_{V}\left(\boldsymbol{Q}\left(t\right)\right)\leq B-V\mathbb{E}\left[D^{*}\left(t\right)|\boldsymbol{Q}\left(t\right)\right]-\sum_{n\in\mathcal{N}}b_{n}Q_{n}\left(t\right)\\
	&+\mathbb{E}\Bigg[\sum_{i\in\mathcal{A}\left(t\right)}\sum_{n\in\mathcal{N}}v_{i,n}(t,\hat{\rho}_{n}(t),\hat{\phi}_{n}(t))\mathds{1}\{I_{i}^{\epsilon}(t)=n\}\Bigg]\\
	&+Va_{\text{max}}\left(\phi_{\text{max}}w_{\text{max}}+\rho_{\text{max}}l_{\text{max}}\right)\\
	&\leq B-\sum_{n\in\mathcal{N}}b_{n}Q_{n}\left(t\right)+Va_{\text{max}}\left(\phi_{\text{max}}w_{\text{max}}+\rho_{\text{max}}l_{\text{max}}\right)\\
	&+\mathbb{E}\Bigg[\sum_{i\in\mathcal{A}\left(t\right)}\sum_{n\in\mathcal{N}}v_{i,n}(t,\hat{\rho}_{n}(t),\hat{\phi}_{n}(t))\mathds{1}\{I_{i}^{\epsilon}(t)=n\}\Bigg].
\end{split}
\end{equation}
The inequality is because $D^{*}(t)\geq 0$.
By the definition of $v_{i,n}(\cdot)$ in (\ref{def: auxiliary func}), it follows that 
\begin{equation}
\begin{split}
	&\Delta_{V}(\boldsymbol{Q}(t))\leq B-\sum_{n\in\mathcal{N}}b_{n}Q_{n}\left(t\right)\\
	&+Va_{\text{max}}\left(\phi_{\text{max}}w_{\text{max}}+\rho_{\text{max}}l_{\text{max}}\right)\\
	&+\sum_{n\in\mathcal{N}}Q_{n}(t)\mathbb{E}\left[E_{n}^{\epsilon}(t)\right]+V\sum_{i\in\mathcal{A}(t)}\mathbb{E}\left[\left(\bar{\phi}_{I_{i}^{\epsilon}(t)}(t)W_{i}\left(t\right)\right.\right.\\
	&~~~~~~~~~~~~~~~~~~~~~~~~~~~\left.\left.+\bar{\rho}_{I_{i}^{\epsilon}\left(t\right)}\left(t\right)L_{i}\left(t\right)\mathds{1}\left\{ n>0\right\} \right)\right]\\
	&\leq B+V\left(\theta_{1}+\theta_{2}\right)+\sum_{n\in\mathcal{N}}Q_{n}\left(t\right)\left(\mathbb{E}\left[E_{n}^{\epsilon}\left(t\right)\right]-b_{n}\right),
\end{split}
\end{equation}
where $\theta_{1} =2a_{\text{max}}\phi_{\text{max}}w_{\text{max}}$ and $\theta_{2}=2a_{\text{max}}\rho_{\text{max}}l_{\text{max}}$. 
It follows by (\ref{ineq: energy constraint under A-only}) that
\begin{equation}
\begin{split}
	\Delta_{V}\left(\boldsymbol{Q}\left(t\right)\right)\leq B+V\left(\theta_{1}+\theta_{2}\right)-\epsilon\sum_{n\in\mathcal{N}}Q_{n}\left(t\right).
\end{split}
\end{equation}
Substituting definitions (\ref{def: Lyapunov drift}) and (\ref{def: drift-plus-penalty}) into above inequality yields
\begin{multline}
	L\left(\boldsymbol{Q}\left(t+1\right)\right)-L\left(\boldsymbol{Q} \left(t\right)\right)+V\mathbb{E}\left[\Delta R\left(t\right)|\boldsymbol{Q}\left(t\right)\right]\\
	\leq B+V\left(\theta_{1}+\theta_{2}\right)-\epsilon\sum_{n\in\mathcal{N}}Q_{n}\left(t\right),
\end{multline}
then by $\Delta R(t)\geq 0$ we have
\begin{multline}
	L\left(\boldsymbol{Q} \left(t+1\right)\right)-L\left(\boldsymbol{Q}\left(t\right)\right)\\
	\leq B+V\left(\theta_{1}+\theta_{2}\right)-\epsilon \sum_{n\in\mathcal{N}}Q_{n}\left(t\right).
\end{multline}
Taking expectation of both sides of the inequality and summing over the first $t$ time slots, we have
\begin{multline}
	\mathbb{E}\left[L\left(\boldsymbol{Q}\left(t\right)\right)\right]-\mathbb{E}\left[L\left(\boldsymbol{Q}\left(0\right)\right)\right]\\
	\leq\left(B+V\left(\theta_{1}+\theta_{2}\right)\right)t-\epsilon\sum_{\tau=0}^{t-1}\sum_{n\in\mathcal{N}}\mathbb{E}\left[Q_{n}\left(\tau\right)\right].
\end{multline}
Dividing both side by $t\epsilon$ and rearrange the items, we have
\begin{multline}
	\frac{1}{t}\sum_{\tau=0}^{t-1}\sum_{n\in\mathcal{N}}\mathbb{E}\left[Q_{n}\left(\tau\right)\right]+\frac{1}{t\epsilon}\mathbb{E}\left[L\left(\boldsymbol{Q}\left(t\right)\right)\right]\\
	\leq \frac{B+V\left(\theta_{1}+\theta_{2}\right)}{\epsilon}+\frac{\mathbb{E}\left[L\left(\boldsymbol{Q}\left(0\right)\right)\right]}{t\epsilon}.
\end{multline}
Since $L(\boldsymbol{Q}(0))=0$ and $L(\boldsymbol{Q} (t))\geq 0$ for all $t$, we have 
\begin{equation}
	\frac{1}{t}\sum_{\tau=0}^{t-1}\sum_{n\in\mathcal{N}}\mathbb{E}\left[Q_{n}\left(\tau\right)\right]\leq \frac{B+V\left(\theta_{1}+\theta_{2}\right)}{\epsilon}.
\end{equation}
Then, letting $t\rightarrow\infty$ gives
\begin{equation}
	\limsup_{t\rightarrow\infty}\frac{1}{t}\sum_{\tau=0}^{t-1}\sum_{n\in\mathcal{N}}\mathbb{E}\left[Q_{n}(\tau)\right]\leq\frac{B+V(\theta_{1}+\theta_{2})}{\epsilon},
\end{equation}
which implies that 
\begin{equation}
	\limsup_{t\rightarrow\infty}\frac{1}{t}\sum_{\tau=0}^{t-1}\mathbb{E}\left[Q_{n}\left(\tau\right)\right]<\infty,\ \forall n\in\mathcal{N}.
\end{equation}

\section{Proof of Theorem \ref{theorem: regret bound}}\label{proof: regret bound}

We consider an optimal $S$-only policy which makes offloading decision $\{I_{i}^{*}(t)\}_{i\in\mathcal{A}(t)}$ in every time slot $t$. According to (\ref{ineq: difference bound 2}) and (\ref{def: one-slot regret}), we have
\begin{equation}\label{ineq: drift-plus-regret bound 2}
\begin{split}
	&L\left(\boldsymbol{Q}\left(t+1\right)\right)-L\left(\boldsymbol{Q}\left(t\right)\right)+V\Delta R\left(t\right)\\
	&\leq B+\sum_{n\in\mathcal{N}}Q_{n}\left(t\right)\left(E_{n}\left(t\right)-b_{n}\right)\\
	&+V\sum_{i\in\mathcal{A}(t)}\left(\rho_{I_{i}(t)}L_{i}(t)\mathds{1}\left\{ I_{i}(t)>0\right\} +\phi_{I_{i}(t)}W_{i}(t)\right)\\
	&-V\sum_{i\in\mathcal{A}(t)}\left(\rho_{I_{i}^{*}(t)}L_{i}(t)\mathds{1}\left\{ I_{i}^{*}(t)>0\right\} +\phi_{I_{i}^{*}(t)}W_{i}(t)\right).
\end{split}
\end{equation}
Substituting (\ref{eq: user energy}) and (\ref{eq: fog energy}) into (\ref{ineq: drift-plus-regret bound 2}) gives
\begin{equation}\label{ineq: difference and regret}
\begin{split}
	&L\left(\boldsymbol{Q}\left(t+1\right)\right)-L\left(\boldsymbol{Q}\left(t\right)\right)+V\Delta R\left(t\right)\leq B\\
	&+\sum_{i\in\mathcal{A}\left(t\right)}\big(Q_{0}\left(t\right)\kappa_{0}\left(t\right)W_{i}\left(t\right)+V\phi_{0}W_{i}\left(t\right)\big)\\
	&~~~~~~~~~~~~~~~~~~~~~ \cdot \left(\mathds{1}\left\{ I_{i}\left(t\right)=0\right\} -\mathds{1}\left\{ I_{i}^{*}\left(t\right)=0\right\} \right)\\
	&+\sum_{i\in\mathcal{A}\left(t\right)}\sum_{n=1}^{N}\big(Q_{n}(t)\kappa_{n}(t)W_{i}(t)+Q_{0}(t)\eta_{n}(t)L_{i}(t)\\
	&~~~~~~~~~~~~~~~~+V\phi_{n}W_{i}\left(t\right)+V\rho_{n}L_{i}\left(t\right)\big)\\
	&~~~~~~~~~~~~~~~~~~~~~\cdot \left(\mathds{1}\left\{ I_{i}\left(t\right)=n\right\} -\mathds{1}\left\{ I_{i}^{*}\left(t\right)=n\right\} \right)\\
	&+Q_{0}(t)\Bigg(\sum_{i\in\mathcal{A}\left(t\right)}\kappa_{0}\left(t\right)W_{i}\left(t\right)\mathds{1}\left\{ I_{i}^{*}\left(t\right)=0\right\}\\
	&~~~~~~~~~+\sum_{i\in\mathcal{A}(t)}\sum_{n=1}^{N}\eta_{n}(t)L_{i}(t)\mathds{1}\{ I_{i}^{*}(t)=n\} -b_{0}\Bigg)\\
	&+\sum_{n=1}^{N}Q_{n}(t)\!\Bigg(\!\sum_{i\in\mathcal{A}(t)}\kappa_{n}(t)W_{i}(t)\mathds{1}\{ I_{i}^{*}(t)=n\} -b_{n}\Bigg).
\end{split}
\end{equation}
Since the optimal policy $\{I_{i}^{*}(t)\}_{i,t}$ is feasible and \textit{i.i.d.} over time, we have
\begin{multline}\label{ineq: optimal user energy}
	\mathbb{E}\Bigg[\sum_{i\in\mathcal{A}\left(t\right)}\kappa_{0}\left(t\right)W_{i}\left(t\right)\mathds{1}\left\{ I_{i}^{*}\left(t\right)=0\right\} \\
	+\sum_{i\in\mathcal{A}\left(t\right)}\sum_{n=1}^{N}\eta_{n}\left(t\right)L_{i}\left(t\right)\mathds{1}\left\{ I_{i}^{*}\left(t\right)=n\right\} \Bigg]\leq b_{0}
\end{multline}
and 
\begin{multline}\label{ineq: optimal fog energy}
	\mathbb{E}\Bigg[\sum_{i\in\mathcal{A}\left(t\right)}\kappa_{n}\left(t\right)W_{i}\left(t\right)\mathds{1}\left\{ I_{i}^{*}\left(t\right)=n\right\} \Bigg]\leq b_{n},\\
	\forall n\in\{1,\dots,N\},
\end{multline}
\textit{i.e.}, the average energy consumption on each node $n\in\mathcal{N}$ in every time slot is no larger than $b_{n}$. Combining (\ref{ineq: difference and regret}), (\ref{ineq: optimal user energy}), and (\ref{ineq: optimal fog energy}) gives
\begin{equation}
\begin{split}
	&\mathbb{E}\left[L\left(\boldsymbol{Q}\left(t+1\right)\right)-L\left(\boldsymbol{Q}\left(t\right)\right)+V\Delta R\left(t\right)\right]\leq B\\
	&+\mathbb{E}\Bigg[\sum_{i\in\mathcal{A}\left(t\right)}\left(Q_{0}\left(t\right)\kappa_{0}\left(t\right)W_{i}\left(t\right)+V\phi_{0}W_{i}\left(t\right)\right)\\
	&~~~~~~~~~~~~~~~~~~~\cdot\left(\mathds{1}\left\{ I_{i}\left(t\right)=0\right\} -\mathds{1}\left\{ I_{i}^{*}\left(t\right)=0\right\} \right)\big]\\
	&+\mathbb{E}\Bigg[\sum_{i\in\mathcal{A}\left(t\right)}\sum_{n=1}^{N}\big(Q_{n}\left(t\right)\kappa_{n}\left(t\right)W_{i}\left(t\right)\\
	&~~~~~+Q_{0}\left(t\right)\eta_{n}\left(t\right)L_{i}\left(t\right)+V\phi_{n}W_{i}\left(t\right)+V\rho_{n}L_{i}\left(t\right)\big)\\
	&~~~~~~~~~~~~~~~~~~~\cdot\left(\mathds{1}\left\{ I_{i}\left(t\right)=n\right\} -\mathds{1}\left\{ I_{i}^{*}\left(t\right)=n\right\} \right)\big].
\end{split}
\end{equation}
By the definition of $v_{i,n}(\cdot)$ in (\ref{def: auxiliary func}), we can rewrite the above inequality as
\begin{equation}
\begin{split}
	&\mathbb{E}\left[L\left(\boldsymbol{Q}\left(t+1\right)\right)-L\left(\boldsymbol{Q}\left(t\right)\right)+V\Delta R\left(t\right)\right]\leq B\\
	&+\mathbb{E}\Bigg[\sum_{i\in\mathcal{A}\left(t\right)}\sum_{n\in\mathcal{N}}v_{i,n}\left(t,\rho_{n},\phi_{n}\right)\\
	&~~~~~~~~~~~~~~~~~~~~\cdot\left(\mathds{1}\left\{ I_{i}\left(t\right)=n\right\}-\mathds{1}\left\{ I_{i}^{*}\left(t\right)=n\right\} \right)\big].
\end{split}
\end{equation}
Define 
\begin{multline}
	C_{1}(t)\triangleq \sum_{i\in\mathcal{A}\left(t\right)}\sum_{n\in\mathcal{N}}v_{i,n}\left(t,\rho_{n},\phi_{n}\right)\\
	\cdot\left(\mathds{1}\left\{ I_{i}\left(t\right)=n\right\} -\mathds{1}\left\{ I_{i}^{*}\left(t\right)=n\right\} \right),
\end{multline}
then 
\begin{multline}\label{ineq: drift-plus-regret bound}
	\mathbb{E}\left[L\left(\boldsymbol{Q}\left(t+1\right)\right)-L\left(\boldsymbol{Q}\left(t\right)\right)+V\Delta R\left(t\right)\right]\\
	\leq B+\mathbb{E}\left[C_{1}\left(t\right)\right].
\end{multline}
Summing (\ref{ineq: drift-plus-regret bound}) over the first $T$ time slots and dividing both sides by $TV$, we obtain
\begin{multline}
	\frac{\mathbb{E}\left[L\left(\boldsymbol{Q}\left(T\right)\right)\right]}{TV}-\frac{\mathbb{E}\left[L\left(\boldsymbol{Q}\left(0\right)\right)\right]}{TV}+\frac{1}{T}\sum_{t=0}^{T-1}\mathbb{E}\left[\Delta R\left(t\right)\right]\\
	\leq\frac{B}{V}+\frac{1}{TV}\sum_{t=0}^{T-1}\mathbb{E}\left[C_{1}\left(t\right)\right].
\end{multline}
Since $L\left(\boldsymbol{Q}\left(0\right)\right)=0$ and $L\left(\boldsymbol{Q}\left(T\right)\right)\geq 0$, we have
\begin{equation}\label{ineq: regret bound}
	\frac{1}{T}\sum_{t=0}^{T-1}\mathbb{E}\left[\Delta R\left(t\right)\right]\leq\frac{B}{V}+\frac{1}{TV}\sum_{t=0}^{T-1}\mathbb{E}\left[C_{1}\left(t\right)\right].
\end{equation}

\subsection{Upper Bound of $C_{1}(t)$}

In this section, we will derive the upper bound of $\mathbb{E}[C_{1}(t)]$.
Consider a policy which makes offloading decision $\{I'_{i}(t)\}_{i\in\mathcal{A}(t)}$ in each time slot $t$ such that
\begin{equation}
	I_{i}'\left(t\right)=\arg\min_{n\in\mathcal{N}\left(t\right)}v_{i,n}\left(t,\rho_{n},\phi_{n}\right),\ \forall i\in\mathcal{A}(t).
\end{equation}
It follows that
\begin{equation}
\begin{split}
	&\sum_{n\in\mathcal{N}}v_{i,n}(t,\rho_{n},\phi_{n})\mathds{1}\{ I'_{i}(t)=n\} \\
	&\leq\sum_{n\in\mathcal{N}}v_{i,n}(t,\rho_{n},\phi_{n})\mathds{1}\{ I^{*}_{i}(t)=n\},\ \forall i\in\mathcal{A}(t).
\end{split}
\end{equation}
Then under our policy LAGO which makes offloading decision $\{I_{i}(t)\}_{i\in\mathcal{A}(t)}$ in each time slot $t$, $C_{1}(t)$ can be upper bounded as follows:
\begin{equation}\label{ineq: C1 bound 1}
\begin{split}
	&C_{1}(t)=\sum_{i\in\mathcal{A}(t)}\sum_{n\in\mathcal{N}}v_{i,n}(t,\rho_{n},\phi_{n})\\
	&~~~~~~~~~~~~~~~~~~~~~~~\cdot\left(\mathds{1}\left\{ I_{i}\left(t\right)=n\right\}-\mathds{1}\{ I_{i}^{*}(t)=n\} \right)\\
	&\leq \sum_{i\in\mathcal{A}\left(t\right)}\sum_{n\in\mathcal{N}}v_{i,n}\left(t,\rho_{n},\phi_{n}\right)\\
	&~~~~~~~~~~~~~~~~~\cdot\left(\mathds{1}\left\{ I_{i}\left(t\right)=n\right\} -\mathds{1}\left\{ I_{i}'\left(t\right)=n\right\} \right)\\
	&\leq \sum_{i\in\mathcal{A}\left(t\right)}\sum_{n\in\mathcal{N}}v_{i,n}\left(t,\rho_{n},\phi_{n}\right)\\
	&~~~~~~~~~~~~~~~~~\cdot\left(\mathds{1}\left\{ I_{i}\left(t\right)=n\right\} -\mathds{1}\left\{ I_{i}'\left(t\right)=n\right\} \right)\\
	&+\sum_{i\in\mathcal{A}\left(t\right)}\sum_{n\in\mathcal{N}}v_{i,n}(t,\hat{\rho}_{n}(t),\hat{\phi}_{n}(t))\\
	&~~~~~~~~~~~~~~~~~\cdot\left(\mathds{1}\left\{ I_{i}'\left(t\right)=n\right\} -\mathds{1}\left\{ I_{i}\left(t\right)=n\right\} \right),
\end{split}
\end{equation}
where the last equality is by (\ref{eq: our policy}). Rearranging the right-hand side of (\ref{ineq: C1 bound 1}), we obtain
\begin{equation}\label{ineq: C1 bound 2}
\begin{split}
	&C_{1}(t)\!\leq\!\sum_{i\in\mathcal{A}(t)}\!\sum_{n\in\mathcal{N}}(v_{i,n}(t,\!\rho_{n},\!\phi_{n})\!-\!v_{i,n}(t,\!\hat{\rho}_{n}(t),\!\hat{\phi}_{n}(t)))\\
	&~~~~~~~~~~~~~~~~~~~~~~\cdot\mathds{1}\left\{ I_{i}\left(t\right)=n\right\} \\
	&+\sum_{i\in\mathcal{A}(t)}\sum_{n\in\mathcal{N}}(v_{i,n}(t,\hat{\rho}_{n}(t),\hat{\phi}_{n}(t))-v_{i,n}(t,\rho_{n},\phi_{n}))\\
	&~~~~~~~~~~~~~~~~\cdot\mathds{1}\left\{ I_{i}'(t)=n\right\} .
\end{split}
\end{equation}
By the definition of $v_{i,n}(\cdot)$ in (\ref{def: auxiliary func}), we have
\begin{equation}\label{eq: price difference}
\begin{split}
	&v_{i,n}(t,\rho_{n},\phi_{n})-v_{i,n}(t,\hat{\rho}_{n}(t),\hat{\phi}_{n}(t))\\
	&=V(\phi_{n}-\hat{\phi}_{n}(t))W_{i}(t)\\
	&+V(\rho_{n}-\hat{\rho}_{n}(t))L_{i}(t)\mathds{1}\{ n>0\}, \forall i\in\mathcal{A}(t), n\in\mathcal{N} .
\end{split}
\end{equation}
Plugging (\ref{eq: price difference}) into the right-hand side of (\ref{ineq: C1 bound 2}) gives
\begin{equation}\label{ineq: C1 bound}
\begin{split}
	&C_{1}(t)\leq V\sum_{i\in\mathcal{A}\left(t\right)}\sum_{n\in\mathcal{N}}\big[(\phi_{n}-\hat{\phi}_{n}(t))W_{i}(t)\\
	&~~~~~~~~~~+(\rho_{n}-\hat{\rho}_{n}(t))L_{i}(t)\mathds{1}\{ n>0\} \big]\mathds{1}\{ I_{i}(t)=n\} \\
	&+V\sum_{i\in\mathcal{A}(t)}\sum_{n\in\mathcal{N}}\big[(\hat{\phi}_{n}(t)-\phi_{n})W_{i}(t)\\
	&~~~~~~~~~+(\hat{\rho}_{n}(t)-\rho_{n})L_{i}(t)\mathds{1}\{ n>0\} \big]\mathds{1}\left\{ I_{i}'(t)=n\right\} .
\end{split}
\end{equation}
Define 
\begin{multline}\label{def: C2}
	C_{2}\left(t\right)\triangleq \sum_{i\in\mathcal{A}(t)}\sum_{n\in\mathcal{N}}\big[(\phi_{n}-\hat{\phi}_{n}(t))W_{i}(t)\\
	+(\rho_{n}-\hat{\rho}_{n}(t))L_{i}(t)\mathds{1}\left\{ n>0\right\} \big]\mathds{1}\left\{ I_{i}\left(t\right)=n\right\}  
\end{multline}
and 
\begin{multline}\label{def: C3}
	C_{3}\left(t\right)\triangleq \sum_{i\in\mathcal{A}(t)}\sum_{n\in\mathcal{N}}\big[(\hat{\phi}_{n}(t)-\phi_{n})W_{i}(t)\\
	+(\hat{\rho}_{n}(t)-\rho_{n})L_{i}(t)\mathds{1}\left\{ n>0\right\} \big]\mathds{1}\left\{ I_{i}'\left(t\right)=n\right\},
\end{multline}
then the upper bound of $C_{1}(t)$ in (\ref{ineq: C1 bound}) can be written as follows:
\begin{equation}
	C_{1}\left(t\right)\leq V\left(C_{2}\left(t\right)+C_{3}\left(t\right)\right).
\end{equation}
Taking expectation of both sides and summing over the first $T$ time slots gives
\begin{equation}\label{ineq: C1 bound 1}
	\sum_{t=0}^{T-1}\mathbb{E}[C_{1}\left(t\right)]\leq V\left(\sum_{t=0}^{T-1}\mathbb{E}[C_{2}\left(t\right)]+\sum_{t=0}^{T-1}\mathbb{E}[C_{3}\left(t\right)]\right).
\end{equation}

\subsection{Upper Bound of $C_{2}(t)$}
In this section, we develop the upper bound of $C_{2}(t)$.
Define event $G_{n}(t)\triangleq\{ \phi_{n}<\hat{\phi}_{n}(t)\} $ for each $n\in\mathcal{N}$ and define event $J_{n}(t)\triangleq\{ \rho_{n}<\hat{\rho}_{n}(t)\} $ for each $n\in\mathcal{N}\setminus \{0\}$, then we have
\begin{equation}\label{ineq: C2 bound}
\begin{split}
	&\mathbb{E}[C_{2}(t)]=\sum_{n\in\mathcal{N}}\sum_{i\in\mathcal{A}(t)}\mathbb{E}\big[(\phi_{n}-\hat{\phi}_{n}(t))W_{i}(t)\\\
	&~~~~~~~~~~~~\cdot\left(\mathds{1}\left\{ G_{n}(t)\right\} +\mathds{1}\left\{ G_{n}^{c}(t)\right\} \right)\big]\mathds{1}\left\{ I_{i}(t)=n\right\} \\
	&+\sum_{n\in\mathcal{N}}\sum_{i\in\mathcal{A}(t)}\mathbb{E}\big[\left(\rho_{n}-\hat{\rho}_{n}\left(t\right)\right)L_{i}\left(t\right)\mathds{1}\left\{ n>0\right\}\\
	&~~~~~~~~~~~~~~~\cdot\left(\mathds{1}\{ J_{n}(t)\} +\mathds{1}\{ J_{n}^{c}(t)\} \right)\big]\mathds{1}\{ I_{i}(t)=n\} \\
	&\stackrel{(a)}{\leq}\sum_{n\in\mathcal{N}}\sum_{i\in\mathcal{A}(t)}\mathbb{E}\big[(\phi_{n}-\hat{\phi}_{n}(t))W_{i}(t)\mathds{1}\left\{ G_{n}^{c}(t)\right\} \big]
	\\
	&~~~~~~~~~~~~~~~~~~~~~~~~~~~~~~~~~~~~~~~~~~~~\cdot\mathds{1}\{ I_{i}(t)=n\} \\
	&+\!\sum_{n\in\mathcal{N}}\!\sum_{i\in\mathcal{A}(t)}\!\mathbb{E}\big[(\rho_{n}-\hat{\rho}_{n}(t))L_{i}(t)\mathds{1}\{ n>0\} \mathds{1}\{ J_{n}^{c}(t)\}\! \big]\\
	&~~~~~~~~~~~~~~~~~~~~~~~~~~~~~~~~~~~~~~~~~~~~\cdot\mathds{1}\left\{ I_{i}(t)=n\right\} \\
	&\stackrel{(b)}{\leq} w_{\text{max}}\sum_{n\in\mathcal{N}}\sum_{i\in\mathcal{A}(t)}\mathbb{E}\big[(\phi_{n}-\hat{\phi}_{n}(t))\mathds{1}\{ G_{n}^{c}(t)\}\big]\\
	&~~~~~~~~~~~~~~~~~~~~~~~~~~~~~~~~~~~~~~~~~~~~\cdot\mathds{1}\left\{ I_{i}\left(t\right)=n\right\} \\
	&+l_{\text{max}}\sum_{n\in\mathcal{N}}\sum_{i\in\mathcal{A}(t)}\mathbb{E}\big[(\rho_{n}-\hat{\rho}_{n}(t))\mathds{1}\{ n>0\} \mathds{1}\{ J_{n}^{c}(t)\} \big]\\
	&~~~~~~~~~~~~~~~~~~~~~~~~~~~~~~~~~~~~~~~~~~~\cdot\mathds{1}\left\{ I_{i}\left(t\right)=n\right\} ,
\end{split}
\end{equation}
where inequality (a) is because $(\phi_{n}-\hat{\phi}_{n}(t))\mathds{1}\left\{ G_{n}(t)\right\} \leq0$ and $\left(\rho_{n}-\hat{\rho}_{n}\left(t\right)\right)\mathds{1}\left\{ J_{n}(t)\right\} \leq0$, and inequality (b) is because $W_{i}(t)\leq w_{\text{max}}$ and $L_{i}(t)\leq l_{\text{max}}$ for all $i\in\mathcal{A}(t)$.
Define
\begin{equation}
	C_{4,n}(t)\triangleq\!\! \sum_{i\in\mathcal{A}(t)}(\phi_{n}-\hat{\phi}_{n}(t))\mathds{1}\{ G_{n}^{c}(t)\}\mathds{1}\{ I_{i}(t)=n\}
\end{equation}
for each node $n\in\mathcal{N}$ and define
\begin{multline}
	C_{5,n}(t)\triangleq \sum_{i\in\mathcal{A}(t)}\left(\rho_{n}-\hat{\rho}_{n}(t)\right)\mathds{1}\left\{ n>0\right\} \mathds{1}\left\{ J_{n}^{c}\left(t\right)\right\}\\
	\cdot\mathds{1}\left\{ I_{i}\left(t\right)=n\right\}
\end{multline}
for each fog node $n\in\mathcal{N}\setminus\{0\}$,
then the upper bound of $\mathbb{E}[C_{2}(t)]$ in (\ref{ineq: C2 bound}) can be written as
\begin{multline}\label{ineq: C2 bound 1}
	\mathbb{E}\left[C_{2}\left(t\right)\right]\leq w_{\text{max}}\sum_{n\in\mathcal{N}}\mathbb{E}\left[C_{4,n}\left(t\right)\right]\\
	+l_{\text{max}}\sum_{n\in\mathcal{N}\setminus\left\{ 0\right\} }\mathbb{E}\left[C_{5,n}\left(t\right)\right].
\end{multline}
Summing $\mathbb{E}[C_{2}(t)]$ over the first $T$ time slots gives
\begin{multline}\label{ineq: C2 bound 3}
	\sum_{t=0}^{T-1}\mathbb{E}\left[C_{2}\left(t\right)\right]\leq w_{\text{max}}\sum_{t=0}^{T-1}\sum_{n\in\mathcal{N}}\mathbb{E}\left[C_{4,n}\left(t\right)\right]\\
	+l_{\text{max}}\sum_{t=0}^{T-1}\sum_{n=1}^{N}\mathbb{E}\left[C_{5,n}\left(t\right)\right].
\end{multline}

Let the first time when node $n$ is chosen be time slot $t_{n}$.
Define event $K_{n}\left(t\right)\triangleq\left\{ \phi_{n}-\bar{\phi}_{n}\left(t\right)\leq \phi_{\text{max}}\sqrt{\frac{3\log t}{2h_{n}\left(t\right)}}\right\}$ for each $n\in\mathcal{N}$.
Summing $C_{4,n}(t)$ over the first $T$ time slots gives
\begin{equation}\label{ineq: C4 bound 1}
\begin{split}
	&\sum_{t=0}^{T-1}C_{4,n}\left(t\right)= \sum_{t=0}^{T-1}\sum_{i\in\mathcal{A}(t)}(\phi_{n}-\hat{\phi}_{n}(t))\mathds{1}\left\{ G_{n}^{c}(t)\right\} \\
	&~~~~~~~~~~~~~~~~~~~~~~~~~~~~~~~~~~~~~~~~~~~\cdot\mathds{1}\left\{ I_{i}(t)=n\right\} \\
	&\stackrel{(a)}{=} \sum_{t=t_{n}}^{T-1}\sum_{i\in\mathcal{A}(t)}(\phi_{n}-\hat{\phi}_{n}(t))\mathds{1}\{ G_{n}^{c}(t)\} \mathds{1}\{ I_{i}(t)=n\} \\
	&\stackrel{(b)}{\leq} \phi_{\text{max}}\\
	&+ \sum_{t=t_{n}+1}^{T-1}\sum_{i\in\mathcal{A}(t)}(\phi_{n}-\hat{\phi}_{n}(t))\mathds{1}\{ G_{n}^{c}(t)\} \mathds{1}\{ I_{i}(t)=n\} \\
	&= \phi_{\text{max}}+\sum_{t=t_{n}+1}^{T-1}\sum_{i\in\mathcal{A}(t)}(\phi_{n}-\hat{\phi}_{n}(t))\mathds{1}\{ G_{n}^{c}(t)\} \\
	&~~~~~~~~~~~~~\cdot\left(\mathds{1}\{ K_{n}(t)\}+\mathds{1}\left\{ K_{n}^{c}\left(t\right)\right\} \right)\mathds{1}\left\{ I_{i}\left(t\right)=n\right\} \\
	&= \phi_{\text{max}}+\sum_{t=t_{n}+1}^{T-1}\sum_{i\in\mathcal{A}(t)}(\phi_{n}-\hat{\phi}_{n}(t))\\
	&~~~~~~~~~~~~~~~~~~~~~~~\cdot\mathds{1}\{ G_{n}^{c}(t)\cap K_{n}(t)\}\mathds{1}\{ I_{i}(t)=n\} \\
	&+\sum_{t=t_{n}+1}^{T-1}\sum_{i\in\mathcal{A}(t)}(\phi_{n}-\hat{\phi}_{n}(t))\mathds{1}\{ K_{n}^{c}(t)\} \mathds{1}\{ I_{i}(t)=n\} \\
	&\stackrel{(c)}{\leq}\phi_{\text{max}}+ \sum_{t=t_{n}+1}^{T-1}\sum_{i\in\mathcal{A}(t)}(\phi_{n}-\hat{\phi}_{n}(t))\\
	&~~~~~~~~~~~~~~~~~~~~~~~\cdot\mathds{1}\{ G_{n}^{c}(t)\cap K_{n}(t)\}\mathds{1}\left\{ I_{i}(t)=n\right\} \\
	&+\phi_{\text{max}}\sum_{t=t_{n}+1}^{T-1}\sum_{i\in\mathcal{A}(t)}\mathds{1}\left\{ K_{n}^{c}(t)\right\} \mathds{1}\left\{ I_{i}(t)=n\right\}.
\end{split}
\end{equation}
Equality (a) is because $\mathds{1}\left\{ I_{i}\left(t\right)=n\right\}=0$ when $t< t_{n}$. Inequalities (b) and (c) are because $\phi_{n}\leq \phi_{\text{max}}$ and $\hat{\phi}_{n}(t)\geq 0$.
Define
\begin{multline}\label{eq: U1 definition}
	U_{1,n}(t)\triangleq\sum_{i\in\mathcal{A}(t)}(\phi_{n}-\hat{\phi}_{n}(t))\mathds{1}\left\{ G_{n}^{c}(t)\cap K_{n}(t)\right\}\\
	\cdot\mathds{1}\left\{ I_{i}(t)=n\right\}
\end{multline}
and 
\begin{equation}\label{eq: U2 definition}
\begin{split}
	U_{2,n}\left(t\right)\triangleq\sum_{i\in\mathcal{A}\left(t\right)}\mathds{1}\left\{ K_{n}^{c}\left(t\right)\right\} \mathds{1}\left\{ I_{i}\left(t\right)=n\right\},
\end{split}
\end{equation}
then by (\ref{ineq: C4 bound 1}) we have
\begin{multline}
	\sum_{t=0}^{T-1}\sum_{n\in\mathcal{N}}\mathbb{E}\left[C_{4,n}\left(t\right)\right]\leq \phi_{\text{max}}\\
	+\sum_{t=t_{n}+1}^{T-1}\sum_{n\in\mathcal{N}}\left(\mathbb{E}\left[U_{1,n}\left(t\right)\right]+\phi_{\text{max}}\mathbb{E}\left[U_{2,n}\left(t\right)\right]\right).
\end{multline}

We first bound $U_{i,n}(t)$. Note that when the event $K_{n}\left(t\right)$ happens, we have $\phi_{n}-\bar{\phi}_{n}\left(t\right)\leq \phi_{\text{max}}\sqrt{3\log t/(2h_{n}\left(t\right))}$. By this, along with (\ref{eq: UCB estimate about process}), which implies $\bar{\phi}_{n}\left(t\right)-\hat{\phi}_{n}\left(t\right)\leq\sqrt{3\log t/(2h_{n}\left(t\right))}$, we obtain
\begin{multline}\label{ineq: U1 bound process 1}
	\phi_{n}-\hat{\phi}_{n}\left(t\right)=\left(\phi_{n}-\bar{\phi}_{n}\left(t\right)\right)+\left(\bar{\phi}_{n}\left(t\right)-\hat{\phi}_{n}\left(t\right)\right)\\
	\leq2\phi_{\text{max}}\sqrt{\frac{3\log t}{2h_{n}\left(t\right)}}.
\end{multline}
Plugging (\ref{ineq: U1 bound process 1}) into (\ref{eq: U1 definition}) and then summing over the time slots $\{t_{n}+1,\dots,T-1\}$ yields
\begin{equation}
\begin{split}
	&\sum_{t=t_{n}+1}^{T-1}U_{1,n}\left(t\right)=\sum_{t=t_{n}+1}^{T-1} \sum_{i\in\mathcal{A}\left(t\right)}(\phi_{n}-\hat{\phi}_{n}\left(t\right))\\
	&~~~~~~~~~~~~~~~~~~~~~~\cdot\mathds{1}\left\{ G_{n}^{c}(t)\cap K_{n}(t)\right\} \mathds{1}\left\{ I_{i}\left(t\right)=n\right\} \\
	&\leq \sum_{t=t_{n}+1}^{T-1}\sum_{i\in\mathcal{A}(t)}2\phi_{\text{max}}\sqrt{\frac{3\log t}{2h_{n}(t)}}\mathds{1}\left\{ G_{n}^{c}(t)\cap K_{n}(t)\right\} \\
	&~~~~~~~~~~~~~~~~~~~~~~~~~~~~~~~~~~~~~~~~~~~~~\cdot\mathds{1}\left\{ I_{i}\left(t\right)=n\right\} \\
	&\leq \sum_{t=t_{n}+1}^{T-1}\sum_{i\in\mathcal{A}\left(t\right)}2\phi_{\text{max}}\sqrt{\frac{3\log t}{2h_{n}\left(t\right)}}\mathds{1}\left\{ I_{i}\left(t\right)=n\right\} \\
	&\leq \sum_{t=t_{n}+1}^{T-1}\sum_{i\in\mathcal{A}\left(t\right)}2\phi_{\text{max}}\sqrt{\frac{3\log T}{2h_{n}\left(t\right)}}\mathds{1}\left\{ I_{i}\left(t\right)=n\right\} \\
	&= \phi_{\text{max}}\sqrt{6\log T}\sum_{t=t_{n}+1}^{T-1}\frac{1}{\sqrt{h_{n}(t)}}\sum_{i\in\mathcal{A}(t)}\mathds{1}\left\{ I_{i}(t)=n\right\} .
\end{split}
\end{equation}
Let $a_{n}(t)$ be the number of times that node $n$ is chosen in time slot $t$ under our policy LAGO, \textit{i.e.} $a_{n}\left(t\right)=\sum_{i\in\mathcal{A}\left(t\right)}I_{i}\left(t\right)$, then we have
\begin{equation}\label{ineq: U1 bound process 3}
\begin{split}
	\sum_{t=t_{n}+1}^{T-1}U_{1,n}\left(t\right)\leq \phi_{\text{max}}\sqrt{6\log T}\sum_{t=t_{n}+1}^{T-1}\frac{a_{n}\left(t\right)}{\sqrt{h_{n}\left(t\right)}}.
\end{split}
\end{equation}
Let $t_{n,m}$ be the $m$th time slot when node $n$ is chosen, and let $M_{n}(T)$ be the time slot when node $n$ is lastly chosen in the first $T$ time slots. Then we have
\begin{equation}\label{ineq: U1 bound process 2}
\begin{split}
	&\sum_{t=t_{n}+1}^{T-1}\frac{a_{n}(t)}{\sqrt{h_{n}(t)}}=\sum_{m=2}^{M_{n}(T)}\frac{a_{n}(t_{n,m})}{\sqrt{h_{n}(t_{n,m})}}\\
	&\stackrel{(a)}{\leq}\sum_{m=2}^{h_{n}(T)}\frac{a_{\text{max}}}{\sqrt{m-1}}
	\stackrel{(b)}{\leq} a_{\text{max}}\left(1+\int_{1}^{h_{n}\left(T\right)}\frac{1}{\sqrt{m}}dm\right)\\
	&\leq2a_{\text{max}}\sqrt{h_{n}\left(T\right)}.
\end{split}
\end{equation}
Inequality (a) is because $a_{n}(t)\leq a_{\text{max}}$, $h_{n}(t_{n,m})\geq m-1$, and $M_{n}(T)\leq h_{n}(T)$. Inequality (b) is because of the basic relationship between summation and integral. Plugging (\ref{ineq: U1 bound process 2}) into (\ref{ineq: U1 bound process 3}) yeilds
\begin{equation}\label{ineq: U1 bound process 4}
\begin{split}
	\sum_{t=t_{n}+1}^{T-1}U_{1,n}\left(t\right)\leq 2a_{\text{max}}\phi_{\text{max}}\sqrt{6h_{n}\left(T\right)\log T}.
\end{split}
\end{equation}
By Jensen's inequality, we have 
\begin{multline}
\frac{1}{N+1}\sum_{n\in\mathcal{N}}\sqrt{h_{n}\left(T\right)}\leq\sqrt{\frac{1}{N+1}\sum_{n\in\mathcal{N}}h_{n}\left(T\right)}\\
\leq\sqrt{\frac{1}{N+1}\left(a_{\text{max}}T\right)}.
\end{multline}
Then it follows that
\begin{multline}\label{ineq: C4sub1 bound}
	\sum_{t=t_{n}+1}^{T-1}\sum_{n\in\mathcal{N}}\mathbb{E}\left[U_{1,n}\left(t\right)\right]\\
	\leq 2a_{\text{max}}\phi_{\text{max}}\sqrt{6a_{\text{max}}\left(N+1\right)T\log T}.
\end{multline}

Next, we bound $U_{2,n}(t)$. Adopting the Chernoff-Hoeffding bound we have
\begin{equation}\label{ineq: C4sub2 bound}
\begin{split}
	&\sum_{t=t_{n}+1}^{T-1}\sum_{n\in\mathcal{N}}\mathbb{E}\left[U_{2,n}\left(t\right)\right]\\
	&=\sum_{t=t_{n}+1}^{T-1}\sum_{n\in\mathcal{N}}\sum_{i\in\mathcal{A}\left(t\right)}\mathbb{E}\left[\mathds{1}\left\{ K_{n}^{c}\left(t\right)\right\} \right]\mathds{1}\left\{ I_{i}\left(t\right)=n\right\} \\
	&=\sum_{t=t_{n}+1}^{T-1}\sum_{n\in\mathcal{N}}a_{n}\left(t\right)\mathbb{E}\left[\mathds{1}\left\{ K_{n}^{c}\left(t\right)\right\} \right]\\
	&=\sum_{t=t_{n}+1}^{T-1}\sum_{n\in\mathcal{N}}a_{n}\left(t\right)\Pr\left\{ K_{n}^{c}\left(t\right)\right\} \\
	&=\sum_{t=t_{n}+1}^{T-1}\!\sum_{n\in\mathcal{N}}a_{n}(t)\Pr\left\{ \phi_{n}-\bar{\phi}_{n}(t)\!>\!\phi_{\text{max}}\sqrt{\frac{3\log t}{2h_{n}(t)}}\right\} \\
	&\leq\sum_{t=t_{n}+1}^{T-1}\sum_{n\in\mathcal{N}}a_{n}(t)\exp\left(-\frac{2(h_{n}(t))^{2}}{h_{i}(t)\phi_{\text{max}}^{2}}\cdot \phi_{\text{max}}^{2}\frac{3\log t}{2h_{n}(t)}\right) \\
	&=\sum_{t=t_{n}+1}^{T-1}\sum_{n\in\mathcal{N}}a_{n}\left(t\right)\exp\left(-3\log t\right) \\
	&\leq\sum_{t=1}^{T-1}\sum_{n\in\mathcal{N}}a_{n}\left(t\right)t^{-3} \\
	&\leq a_{\text{max}}\sum_{t=1}^{\infty}t^{-3} =a_{\text{max}}\left(1+\sum_{t=2}^{\infty}t^{-3}\right)\\
	&\leq a_{\text{max}}\left(1+\int_{1}^{\infty}t^{-3}dt\right)=\frac{3}{2}a_{\text{max}}.
\end{split}
\end{equation}
Taking expectation of both sides of (\ref{ineq: C4 bound 1}) and then plugging (\ref{ineq: C4sub1 bound}) (\ref{ineq: C4sub2 bound}) into it yields
\begin{multline}\label{ineq: C4 bound 2}
	\sum_{t=0}^{T-1}\sum_{n\in\mathcal{N}}\mathbb{E}\left[C_{4,n}\left(t\right)\right]\\
	\leq 2a_{\text{max}}\phi_{\text{max}}\sqrt{6a_{\text{max}}\left(N+1\right)T\log T}+\frac{3}{2}a_{\text{max}}\phi_{\text{max}}.
\end{multline}
Similarly, we have
\begin{multline}\label{ineq: C5 bound}
	\sum_{t=0}^{T-1}\sum_{n\in\mathcal{N}}\mathbb{E}\left[C_{5,n}\left(t\right)\right]\\
	\leq 2a_{\text{max}}\rho_{\text{max}}\sqrt{6a_{\text{max}}NT\log T}+\frac{3}{2}a_{\text{max}}\rho_{\text{max}}.
\end{multline}
Plugging (\ref{ineq: C4 bound 2}) and (\ref{ineq: C5 bound}) into (\ref{ineq: C2 bound 3}) yields
\begin{equation}\label{ineq: C2 bound 2}
\begin{split}
	&\sum_{t=0}^{T-1}\mathbb{E}\left[C_{2}\left(t\right)\right]\\
	&\leq w_{\text{max}}a_{\text{max}}\phi_{\text{max}}\left(2\sqrt{6a_{\text{max}}\left(N+1\right)T\log T}+\frac{3}{2}\right)\\
	&+l_{\text{max}}a_{\text{max}}\rho_{\text{max}}\left(2\sqrt{6a_{\text{max}}NT\log T}+\frac{3}{2}\right).
\end{split}
\end{equation}

\subsection{Upper Bound of $C_{3}(t)$}
In this section we bound $C_{3}(t)$.
Recall from (\ref{def: C3}) that 
\begin{multline}
	C_{3}\left(t\right)=\sum_{i\in\mathcal{A}\left(t\right)}\sum_{n\in\mathcal{N}}\big[(\hat{\phi}_{n}\left(t\right)-\phi_{n})W_{i}\left(t\right)\\
	+\left(\hat{\rho}_{n}\left(t\right)-\rho_{n}\right)L_{i}\left(t\right)\mathds{1}\left\{ n>0\right\} \big]\mathds{1}\left\{ I_{i}'\left(t\right)=n\right\} .
\end{multline}
Taking expectation of the both sides gives
\begin{equation}\label{ineq: C3 Bound 1}
\begin{split}
	&\mathbb{E}\left[C_{3}\left(t\right)\right]=\mathbb{E}\Bigg[\sum_{i\in\mathcal{A}\left(t\right)}\sum_{n\in\mathcal{N}}\big[(\hat{\phi}_{n}\left(t\right)-\phi_{n})W_{i}\left(t\right)\\
	&~~~~~~~~~\cdot\left(\mathds{1}\left\{ G_{n}\left(t\right)\right\} +\mathds{1}\left\{ G_{n}^{c}\left(t\right)\right\} \right)\big]\mathds{1}\left\{ I_{i}'\left(t\right)=n\right\} \Big]\\
	&+\mathbb{E}\Bigg[\sum_{i\in\mathcal{A}\left(t\right)}\sum_{n\in\mathcal{N}}\big[\left(\hat{\rho}_{n}\left(t\right)-\rho_{n}\right)L_{i}\left(t\right)\mathds{1}\left\{ n>0\right\} \\
	&~~~~~~~~~~\cdot\left(\mathds{1}\left\{ J_{n}\left(t\right)\right\} +\mathds{1}\left\{ J_{n}^{c}\left(t\right)\right\} \right)\big]\mathds{1}\left\{ I_{i}'\left(t\right)=n\right\} \Big]\\
	&\leq \mathbb{E}\Bigg[\sum_{i\in\mathcal{A}\left(t\right)}\sum_{n\in\mathcal{N}}\big[(\hat{\phi}_{n}(t)-\phi_{n})W_{i}(t)\mathds{1}\left\{ G_{n}\left(t\right)\right\}\big]\\
	&~~~~~~~~~~~~~~~~~~~~~~~~~~~~~~~~~~~~~~~~~~ \cdot\mathds{1}\left\{ I_{i}'\left(t\right)=n\right\} \Big]\\
	&+\mathbb{E}\Bigg[\sum_{i\in\mathcal{A}(t)}\sum_{n\in\mathcal{N}}\big[(\hat{\rho}_{n}(t)-\rho_{n})L_{i}(t)\mathds{1}\left\{ J_{n}(t)\right\} \big] \\
	&~~~~~~~~~~~~~~~~~~~~~~~~~~~~~~\cdot \mathds{1}\left\{ n>0\right\}\mathds{1}\left\{ I_{i}'\left(t\right)=n\right\} \Big] \\
	&\leq w_{\text{max}}\mathbb{E}\Bigg[\sum_{i\in\mathcal{A}(t)}\sum_{n\in\mathcal{N}}\big[(\hat{\phi}_{n}(t)-\phi_{n})\mathds{1}\left\{ G_{n}(t)\right\} \big]\\
	&~~~~~~~~~~~~~~~~~~~~~~~~~~~~~~~~~~~~~~~~~~~\cdot\mathds{1}\left\{ I_{i}'\left(t\right)=n\right\} \Big] \\
	&+\!l_{\text{max}}\mathbb{E}\Bigg[\!\sum_{i\in\mathcal{A}(t)}\!\sum_{n\in\mathcal{N}}\big[(\hat{\rho}_{n}(t)\!-\!\rho_{n})\mathds{1}\{ n>0\} \mathds{1}\left\{ J_{n}(t)\right\} \big]\\
	&~~~~~~~~~~~~~~~~~~~~~~~~~~~~~~~~~~~~~~~~~~~\cdot\mathds{1}\left\{ I_{i}'(t)=n\right\} \Big].
\end{split}
\end{equation}
We define 
\begin{equation}\label{def: C6}
	C_{6,n}(t)\triangleq \sum_{i\in\mathcal{A}(t)}[(\hat{\phi}_{n}(t)-\phi_{n})\mathds{1}\{ G_{n}(t)\}]\mathds{1}\{ I_{i}'(t)=n\} 
\end{equation}
for each $n\in\mathcal{N}$, and define 
\begin{multline}\label{def: C7}
	C_{7,n}(t)\triangleq \sum_{i\in\mathcal{A}(t)}\left[(\hat{\rho}_{n}(t)-\rho_{n})\mathds{1}\{ n>0\} \mathds{1}\left\{ J_{n}\left(t\right)\right\} \right]\\
	\cdot\mathds{1}\left\{ I_{i}'\left(t\right)=n\right\} 
\end{multline}
for each $n\in\mathcal{N}\setminus\{0\}$. Then the upper bound of $\mathbb{E}[C_{3}(t)]$ can be written as
\begin{multline}
	\mathbb{E}\left[C_{3}\left(t\right)\right]\leq w_{\text{max}}\sum_{n\in\mathcal{N}}\mathbb{E}\left[C_{6,n}\left(t\right)\right]\\
	+l_{\text{max}}\sum_{n=1}^{N}\mathbb{E}\left[C_{7,n}\left(t\right)\right].
\end{multline}
Summing $\mathbb{E}[C_{3}(t)]$ over the first $T$ time slots yields
\begin{multline}\label{ineq: C3 Bound}
	\sum_{t=0}^{T-1}\mathbb{E}\left[C_{3}\left(t\right)\right]\leq w_{\text{max}}\sum_{t=0}^{T-1}\sum_{n\in\mathcal{N}}\mathbb{E}\left[C_{6,n}\left(t\right)\right]\\
	+l_{\text{max}}\sum_{t=0}^{T-1}\sum_{n=1}^{N}\mathbb{E}\left[C_{7,n}\left(t\right)\right].
\end{multline}

We consider the case when $t\leq t_{n}$ and $t\geq t_{n}+1$ separately, where $t_{n}$ is the first time slot when node $n$ is chosen.
When $t\leq t_{n}$, $\hat{\phi}_{n}(t)=0$ and the event $G_{n}\left(t\right)=\{ \phi_{n}<\hat{\phi}_{n}\left(t\right)\} $ will not occur. Thus $C_{6,n}(t)=0$ when $t\leq t_{n}$.

When $t\geq t_{n}+1$, suppose event $G_{n}(t)$ occurs. Then we have $\hat{\phi}_{n}(t)> \phi_{n}\geq 0$, which implies that $\hat{\phi}_{n}=\bar{\phi}_{n}\left(t\right)-\phi_{\text{max}}\sqrt{\frac{3\log t}{2h_{n}\left(t\right)}}>0$ and it follows that $\phi_{n}<\bar{\phi}_{n}\left(t\right)-\phi_{\text{max}}\sqrt{\frac{3\log t}{2h_{n}\left(t\right)}}$. Thus we can bound $\mathbb{E}[C_{6,n}(t)]$ as follows:
\begin{equation}\label{ineq: C6 bound 3}
\begin{split}
	&\mathbb{E}\left[C_{6,n}\left(t\right)\right]\\
	&=\sum_{i\in\mathcal{A}\left(t\right)}\mathbb{E}\big[(\hat{\phi}_{n}\left(t\right)-\phi_{n})\mathds{1}\left\{ G_{n}\left(t\right)\right\} \big]\mathds{1}\{ I_{i}'(t)=n\} \\
	&\leq \phi_{\text{max}}\sum_{i\in\mathcal{A}\left(t\right)}\mathbb{E}\left[\mathds{1}\left\{ G_{n}\left(t\right)\right\} \right]\mathds{1}\left\{ I_{i}'\left(t\right)=n\right\} \\
	&= \phi_{\text{max}}\sum_{i\in\mathcal{A}\left(t\right)}\Pr\left\{ G_{n}\left(t\right)\right\} \mathds{1}\left\{ I_{i}'\left(t\right)=n\right\} \\
	&=\phi_{\text{max}}\sum_{i\in\mathcal{A}(t)}\Pr\left\{ \phi_{n}\leq\bar{\phi}_{n}(t)-\phi_{\text{max}}\sqrt{\frac{3\log t}{2h_{n}\left(t\right)}}\right\} \\
	&~~~~~~~~~~~~~~~~~~~~~~~~~~~~~~~~~~~~~~~~\cdot\mathds{1}\left\{ I_{i}'\left(t\right)=n\right\} .
\end{split}
\end{equation}
By Chernoff-Hoeffding bound, we have
\begin{equation}\label{ineq: C6 bound process}
\begin{split}
	&\Pr\left\{ \phi_{n}\leq\bar{\phi}_{n}\left(t\right)-\phi_{\text{max}}\sqrt{\frac{3\log t}{2h_{n}\left(t\right)}}\right\} \\
	&=\Pr\left\{ \bar{\phi}_{n}\left(t\right)\geq \phi_{n}+\phi_{\text{max}}\sqrt{\frac{3\log t}{2h_{n}\left(t\right)}}\right\} \\
	&\leq\exp\left(-\frac{2\left(h_{n}\left(t\right)\right)^{2}}{h_{n}\left(t\right)\phi_{\text{max}}^{2}}\cdot \phi_{\text{max}}^{2}\frac{3\log t}{2h_{n}\left(t\right)}\right)\\
	&=\exp\left(-3\log t\right)=t^{-3}.
\end{split}
\end{equation}
Plugging (\ref{ineq: C6 bound process}) into (\ref{ineq: C6 bound 3}) gives
\begin{equation}\label{ineq: C6 bound 1}
\begin{split}
	\mathbb{E}\left[C_{6,n}\left(t\right)\right]\leq \phi_{\text{max}}a'_{n}\left(t\right)t^{-3},
\end{split}
\end{equation}
where $a_{n}'(t)$ is the number of times that node $n$ is chosen in time slot $t$ by policy $\{I_{i}'(t)\}_{i}$, \textit{i.e.}, $a_{n}'\left(t\right)=\sum_{i\in\mathcal{A}\left(t\right)}I_{i}'\left(t\right)$. 
Summing $\mathbb{E}[C_{6,n}(t)]$ over time slots $\{0,\dots,T-1\}$ and nodes $n\in\mathcal{N}$, and applying (\ref{ineq: C6 bound 1}), we obtain
\begin{equation}\label{ineq: C6 bound 2}
\begin{split}
	&\sum_{t=0}^{T-1}\sum_{n\in\mathcal{N}}\mathbb{E}\left[C_{6,n}\left(t\right)\right]\leq \phi_{\text{max}}\sum_{n\in\mathcal{N}}\sum_{t=t_{n}+1}^{T-1}a'_{n}\left(t\right)t^{-3}\\
	&\leq \phi_{\text{max}}\sum_{n\in\mathcal{N}}\sum_{t=1}^{T-1}a'_{n}\left(t\right)t^{-3} \\
	&=\phi_{\text{max}}\sum_{t=1}^{T-1}A\left(t\right)t^{-3} \leq \phi_{\text{max}}a_{\text{max}}\sum_{t=1}^{T-1}t^{-3} \\
	&\leq \phi_{\text{max}}a_{\text{max}}\sum_{t=1}^{\infty}t^{-3}=\phi_{\text{max}}a_{\text{max}}\left(1+\sum_{t=2}^{\infty}t^{-3}\right)\\
	&\leq \phi_{\text{max}}a_{\text{max}}\left(1+\int_{1}^{\infty}\frac{1}{t^{3}}dt\right)=\frac{3}{2}\phi_{\text{max}}a_{\text{max}}.
\end{split}
\end{equation}
Similarly, we have
\begin{equation}\label{ineq: C7 bound}
\begin{split}
	\sum_{t=0}^{T-1}\sum_{n=1}^{N}\mathbb{E}\left[C_{7,n}\left(t\right)\right]\leq \frac{3}{2}\rho_{\text{max}}a_{\text{max}}.
\end{split}
\end{equation}

Plugging (\ref{ineq: C6 bound 2}) and (\ref{ineq: C7 bound}) into (\ref{ineq: C3 Bound}) yields
\begin{equation}\label{ineq: C3 bound 2}
\begin{split}
	\sum_{t=0}^{T-1}\mathbb{E}\left[C_{3}\left(t\right)\right]\leq \frac{3}{2}w_{\text{max}}\phi_{\text{max}}a_{\text{max}}+\frac{3}{2}l_{\text{max}}\rho_{\text{max}}a_{\text{max}}.
\end{split}
\end{equation}

\subsection{Upper Bound of Regret}

Plugging (\ref{ineq: C2 bound 2}) and (\ref{ineq: C3 bound 2}) into (\ref{ineq: C1 bound 1}) and then plugging the result into (\ref{ineq: regret bound}) finally gives
\begin{equation}
\begin{split}
	&\frac{1}{T}\sum_{t=0}^{T-1}\mathbb{E}\left[\Delta R\left(t\right)\right]\\
	&\leq \frac{B}{V}+\left(\frac{3}{2T}+\sqrt{\frac{6a_{\text{max}}\left(N+1\right)\log T}{T}}\right)\theta_{1} \\
	&~~~~+\left(\frac{3}{2T}+\sqrt{\frac{6a_{\text{max}}N\log T}{T}}\right)\theta_{2}
\end{split}
\end{equation}
where $B\!\triangleq\!\sum_{n\in\mathcal{N}}b_{n}^{2}/2+a_{\text{max}}^{2}(\max\{ \kappa_{\text{max}}^{2}w_{\text{max}}^{2},\eta_{\text{max}}^{2}l_{\text{max}}^{2}\}/2 +N^{2}a_{\text{max}}^{2}\kappa_{\text{max}}^{2}w_{\text{max}}^{2}/2$, $\theta_{1}\!\triangleq\! 2w_{\text{max}}\phi_{\text{max}}a_{\text{max}}$, $\theta_{2}\!\triangleq \!2l_{\text{max}}\rho_{\text{max}}a_{\text{max}}$.

\end{document}

z